# Lithium and Vanadium Intercalation into Bilayer $V_2Se_2O$: Ferrimagnetic-Ferroelastic Multiferroics and Anomalous and Spin Transport

*Long Zhang, Yuxin Liu, Junfeng Ren, Guangqian Ding, Xiaotian Wang,* Guangxin Ni,* Guoying Gao,* and Zhenxiang Cheng**

L. Zhang, G. Gao

School of Physics and Wuhan National High Magnetic Field Center

Huazhong University of Science and Technology, Wuhan 430074, China

E-mail: guoying_gao@mail.hust.edu.cn

Y. Liu, J. Ren

School of Physics and Electronics, Shandong Provincial Engineering and Technical Center of Light Manipulations & Institute of Materials and Clean Energy

Shandong Normal University, Jinan 250358, China

G. Ding

School of Sciences and Institute for Advanced Sciences

Chongqing University of Posts and Telecommunications, Chongqing 400065, China

X. Wang, Z. Cheng

Institute for Superconducting and Electronic Materials, Faculty of Engineering and Information Sciences

University of Wollongong, Wollongong 2500, Australia

E-mail: xiaotianw@uow.edu.au; cheng@uow.edu.au

G. Ni

Department of Physics

Florida State University, Tallahassee, FL 32306, USA

E-mail: guangxin.ni@magnet.fsu.edu

G. Ni

National High Magnetic Field Laboratory, Tallahassee, FL 32310, USA

Spin splitting in emerging altermagnets is non-relativistic and momentum-dependent, yet energy-independent, and localized in momentum space, posing challenges for practical applications. Here, we propose an intercalation-driven paradigm for altermagnets to attain ameliorative electronic structures, multiferroic characteristics, and anomalous and spin transport functionalities. As a representative system, we investigate electrochemistry- and self-intercalated $V_2Se_2O$ bilayers, building on the recently reported room-temperature K- and Rb-intercalated $V_2Se_2O$ family [*Nat. Phys.* **2025**, *21*, 754; *Nat. Phys.* **2025**, *21*, 760], utilizing density functional theory, Wannier function analyses, Monte Carlo simulations, and non-equilibrium Green's function methods. Intercalation induces room-temperature intralayer ferrimagnetic and interlayer ferromagnetic order (358 K for Li-intercalation and 773 K for V-intercalation), ferroelasticity (~1 % signal intensity), in-plane uniaxial magnetic anisotropy, and metallization, while also modifying the anomalous Hall effect. Notably, Li- and V-intercalated $V_2Se_2O$ bilayers exhibit enhanced spin splitting and half-metallic behavior, respectively, yielding near-perfect spin filtering efficiency. Intercalation substantially enhances spin transport in $V_2Se_2O$-based devices, enabling giant magnetoresistance (877 %), ultra-high thermal tunneling magnetoresistance (~12000 %), and observable spin Seebeck and temperature negative differential resistance effects. This intercalation-driven paradigm expands altermagnetic functionalities through multifunctional integration, offering promising avenues for advanced, miniaturized, room-temperature exploitation of anomalous, electron, and spin transport properties.

## 1. Introduction

The collinear magnetic family encompasses ferromagnetic (FM), antiferromagnetic (AFM), ferrimagnetic (FiM), and the recently emerging altermagnetic (AM) materials.[1-5] Recognized by *Science* as one of the ten biggest breakthroughs in 2024,[6] altermagnets[2] exhibit unique physical and chemical properties, such as non-relativistic, momentum-dependent spin splitting arising from exchange coupling rather than spin-orbit coupling (SOC),[7-9] alongside anomalous Hall effect (AHE),[10-12] and anomalous Nernst effect (ANE),[1,13] garnering significant recent interest.[1] Unlike conventional FM (FiM) spin splitting, AM spin splitting is energy-independent, and it is localized in momentum space. Furthermore, their zero net magnetic moment prevents interference from parasitic magnetic fields, but complicates spin manipulation, data storage and readout. It's noteworthy that altermagnet might acquire additional magnetic/ferroic characteristics, and improved electronic structures through specific approaches, potentially overcoming limitations of current altermagnet research and applications. While often overlooked, FiM materials with small net magnetic moment feature near-AFM/AM high-speed spin dynamics, weak stray fields,[14] as well as energy- and momentum-dependent spin splitting that can be better than intrinsic AM elements. The presence of two inequivalent magnetic sub-lattices enriches further tunability,[5,15,16] making FiMs desirable candidates for non-volatile spin storage and logic devices.

Multiferroics provide an ideal platform for high-density, multifunctional information and energy storage. It has been reported that altermagnets demonstrate ferroelectric/anti-ferroelectric (FE/AFE) coupling with spin in materials such as perovskite $Ca_3Mn_2O_7$, metal-organic framework (MOF) $C(NH_2)_3Cr(HCOO)_3$;[17] transition metal oxides $SrCrO_3$;[18] and phosphorous chalcogenides $CuWP_2S_6$[18] and $MnPSe_3$.[19] Additionally, ferroelastic (FC) materials, characterized by hysteresis between strain and stress, and associated energy storage and release, supply further opportunities when coupled with magnetism.[20,21] Exploring FC features in experiment-feasible above-room-temperature (above-RT) AM-relevant systems is critical for advancing miniaturized spintronics, electronics as well as micro-electro-mechanical systems (MEMSs). Additionally, some multiferroic materials with AM character have also been reported, including $BiFeO_3$,[22,23] $Fe_2Mo_3O_8$,[24] $CuFeS_2$,[25] $MnS_2$,[26] MnSe,[27] MnO,[28] $VOX_2$ (X = Cl, Br, I),[29] $VCl_3$,[30] $Co_2CF_2$,[31] $(CH_3NH_3)_2MnCl_4$,[32] $NH_4Cu(HCOO)_3$,[32] $LiMnO_2$,[33] $MgFe_2N_2$,[34] and others.[35]

For effective spintronic applications, high spin polarization is essential.[36,37] Conventional promising materials for spin transport include half-metals (HMs),[36] bipolar magnetic semiconductors (BMSs),[36,38] unipolar magnetic semiconductors/half-semiconductors (UMSs/HSCs),[39,40] and spin gapless semiconductors (SGSs).[41] We recently demonstrated multistate tunneling electroresistance (TER), tunneling magnetoresistance (TMR), and near-perfect spin filtering effect in above-RT magnetic tunnel junctions (MTJs), composed of AM CrSb electrode, FE $In_2Se_3$ barrier, and FM $Fe_3GaTe_2$ electrode.[8] TMRs modulation by crystal orientation, interface, and layer thickness were reported in AM-based MTJs of $IrO_2/MnF_2/CrO_2$ and $RuO_2/TiO_2/CrO_2$.[42,43] However, all three aforementioned AM-based studies predominantly rely on FM (near-)HM electrodes rather than utilizing AM materials directly,[8,42,43] limiting the exploration of full potential of altermagnets. $MnF_2$ exhibited a far-below-RT Néel temperature ($T_N$) of ~67 K;[44] and the $RuO_2$'s AM character was challenged on account of its proximity to a quantum phase transition[45] or the spin splitting potentially arising from SOC and lattice distortions,[46] restricting practical application. Non-tunable TMR without spin filtering was observed in $Ag/V_2Te_2O/BiOCl/V_2Te_2O/Ag$ MTJ,[12] while theory-designed $Mn_2Se_2O$ based on AM $V_2Se_2O$, served as a FM HM electrode for TMR and spin filtering.[47] Thus, it's urgently essential to manipulate experimental above-RT altermagnets themselves to achieve various magnetic states with high spin polarization and to implement them controllably in spintronic nano-devices, without employing conventional FM (near-)HM electrodes. What's more, the transport performance under temperature gradients, especially at RT, is critical for practical applications, yet that of AM-relevant nano-devices remains largely unexplored.

To address these challenges, we propose an intercalation-driven paradigm in layered AM materials to uncover multifunctional ferroics, tunable electronic structures, and multistate anomalous and spin transport performances. There are numerous layered altermagnets,[1] such as $MnPSe_3$,[19] $Fe_2Se_2O$,[48] and CrO,[49] we select intercalated AM $V_2Se_2O$ as a model system (Figure 1a), in light of very recent experimental realization in RT AM K- and Rb-intercalated forms[50,51] and extensive theoretical and experimental studies of layered $V_2Se_2O$ family.[52-55] Our intercalation-driven paradigm can be applied to the spacious layered altermagnets family for attaining various magnetic and electronic properties, and merits further investigations. Here, two experiment-feasible intercalation methods are considered. Electrochemistry-intercalation, which precisely and reversibly embeds species between layers of electrode materials under controlled voltage or current.[56-58] In the experiment, the alkali metals including K, Rb, and Cs were intercalated into the $V_2Se_2O$ family,[50,51,53-55] and we utilize Li, a similar alkali metal, which demonstrates several superiorities over K, Rb, and Cs: the smaller radius and stronger electronegativity of Li enhance the overall structural stability and reduce distortion and fatigue; the lower standard electrode potential of Li and higher reducibility compared with K, Rb and Cs[57,58] mean electrochemical reactions require less energy and proceed more rapidly. Self-intercalation, copiously employed in experiments for two-dimensional (2D) materials,[59,60] was also considered. This method incorporates native V atoms, avoiding foreign elements and providing better compatibility during growth.[60] In addition to Li- and V-intercalations, the pristine $V_2Se_2O$ bilayer is examined for comparison.

In this work, the intercalation-driven paradigm is exemplified by Li- and V-intercalated $V_2Se_2O$ bilayers, which exhibit multiferroic FiM and FC features, interlayer-transition from AFM to FM, robust above-RT magnetic critical temperatures, tuned magnetic anisotropy, HM/enhanced spin splitting, AHE, multistate thermal giant magnetoresistance (GMR), TMR and spin filtering effect, temperature negative differential resistance (TNDR) effect, and spin Seebeck effect (SSE). The multifunctional integration in electrochemistry- and self-intercalated AM $V_2Se_2O$ pave the way for future miniaturized, energy-efficient, and high-density information processing, sensing, quantum computing, and MEMS technologies. The approach, using intercalation-driven bilayers of a candidate altermagnet, provides a promising route to overcome intrinsic AM limitations and unlock their potential for serviceable employment.

## 2. Results and Discussion

### 2.1. Intercalated Structure of Bilayer $V_2Se_2O$

The optimized lattice constants and atomic distances of Li-electrochemistry-intercalated, V-self-intercalated, and pristine $V_2Se_2O$ bilayers are summarized in Table 1, with the intercalated structure illustrated in Figure 1a. These systems are denoted as Li@$V_2Se_2O$, V@$V_2Se_2O$, and $V_2Se_2O$, respectively. Similar structural distortions have been reported in intercalated $V_2Se_2O$ family.[54,55] The in-plane lattice constant of the pristine bilayer aligns well with previous reports.[61,62] To validate our computational reliability, we also examined the charge density and electronic structure of the $V_2Se_2O$ monolayer (Figure S1), which reveals a band gap of 0.71 eV, consistent with reported 0.68 and 0.72 eV.[52,63]

Following the removal of alkali metals (K, Rb, and Cs),[53-55] neighboring monolayers typically slide by half a lattice constant along in-plane orthogonal directions. As the stacking/sliding effect has been disclosed in the $V_2Se_2O$ family and other AM-relevant materials,[19,53-55] we are targeting the intercalation effect, so sliding isn't involved herein. The stacking of $V_2Se_2O$ layers with and without intercalation is maintained in the same configuration as in the experimental $KV_2Se_2O$ and $Rb_{1-\delta}V_2Te_2O$,[50,51] to isolate the intercalation-induced phenomena and mechanisms from those caused by stacking/sliding effect. The crystal symmetry of the bilayer $V_2Se_2O$ matches that of the experimental bulk counterparts $KV_2Se_2O$, $CsV_2Se_2O$ and $Rb_{1-\delta}V_2Te_2O$, all of which belong to the *P*4/*mmm* (No.123) space group.[50,51,53-55] Upon Li- and V-intercalation, the symmetry lowers to *Pmmm* (No. 47) due to the introduction of an inhomogeneous chemical environment that distorts lattice constants. Similar symmetry reduction after intercalation in experiment has been reported, suggesting our results for symmetry reduction could be feasible.[59,64-66] Specifically, the in-plane lattice parameter $a$ becomes larger than $b$, with the pristine value falling between the two. Intercalation enhances interlayer interactions and reduces the inter-Se-Se distance.

The charge density difference patterns (Figure 2a-f) directly visualize interlayer interactions before and after intercalation. Yellow and blue regions depict charge accumulation and depletion, respectively. Intercalated Li and V atoms donate electrons to neighboring Se atoms, while V atoms within the $V_2Se_2O$ layer show minor charge accumulation. In pristine bilayer $V_2Se_2O$, interlayer electrons are mainly depleted, Se atoms mainly gain electrons and V atoms loss them. Charge transfer has been verified in experimental studies.[50,55,67,68] Scanning tunneling microscope (STM) simulations (Figure 1b-d) provide an intuitive visualization of lattice morphology for pristine and intercalated structures and a reference for experimental growth and processing. Local conductivity variations, particularly around O atoms and central V atoms (elliptical and circular regions) in intercalated bilayers, manifest as color changes and indicate a redistribution of electronic states. These electronic redistributions affect inter-atomic interactions, work functions, and tunneling behavior, which are critical for spintronic and electronic performance, as detailed in subsequent sections.

As shown in Figure 2e-g, the interlayer potential of the pristine bilayer approximates the vacuum level, whereas Li or V intercalation significantly lowers it. Due to their differing electronegativities, Li and V intercalations result in distinct internal potentials within the structures: -2.497 eV for Li and -7.425 eV for V. Intercalation also induces asymmetry between the potentials of the inner and outer Se atoms. Consequently, built-in electric field is directed: (1) from Li to Se, (2) from Se to the intercalated V, and (3) from Se to V and O within the layer. In $V_2Se_2O$ bilayers with Li/V intercalation, the averaged electric field from outer Se to the O and V plane is 9.953/9.453 eV $Å^{-1}$, 8.733/7.652 eV $Å^{-1}$ from inner Se to the O and V planes, and 1.812/0.756 eV $Å^{-1}$ from intercalated Li to inner Se/from inner Se to intercalated V. For pristine $V_2Se_2O$ bilayer, the averaged electric field from outer Se to the O and V planes is 9.757 eV $Å^{-1}$, and from inner Se to the O and V planes is 9.971 eV $Å^{-1}$. Our averaged electric field values (Figure 2e-g) exceed those of typical 2D materials,[3,69] suggesting superior resilience to external electric fields and minimal susceptibility to noise. In all cases, the $V_2Se_2O$ bilayers exhibit mirror symmetry in their electrostatic potential, confirming the absence of ferroelectricity.

Electron localization functions (ELFs) in the (110) plane (Figure 2h-j) indicate that electrons are mostly localized around atomic sites, with bonding characterized as predominantly ionic with minor covalent character. Green regions between atoms show electron distribution, reflecting enhanced conductivity after intercalation.

## 2.2. Robust above-Room-Temperature Magnetism

### 2.2.1. Magnetic Configuration

Stable magnetism in low-dimensional systems is essential for magnetic transport and storage applications. The experimental investigations into $KV_2Se_2O$, $CsV_2Se_2O$, and $Rb_{1-\delta}V_2Te_2O$ have unveiled that these intercalated $V_2Se_2O$-family materials preserve a collinear spin alignment.[50,51,53-55] These findings suggest that the magnetic configuration of intercalated $V_2Se_2O$ systems in our study appears to lean toward a collinear arrangement. The DMI is forbidden for $Cr_2Se_2O$[70] based on the Moriya symmetry rules.[71] The non-collinear spin alignment and Dzyaloshinskii-Moriya interaction (DMI) necessitate the breaking of spatial symmetry, such as Janus $V_2SeTeO$ monolayer.[72] However, the intercalated $V_2Se_2O$ systems retain their spatial symmetry. Considering intralayer and interlayer collinear magnetic configurations (Figure S2) based on the prior studies,[12,52] the calculated energy differences among these states, relative to the intralayer AFM1 and interlayer FM (AFM1-FM) state, are displayed in Figure 3a. The anisotropy energy differences according to the magnetic ground states (Figure 3b) will be discussed later. It's evident that the magnetic ground states of Li- and V-intercalated $V_2Se_2O$ bilayers favor intralayer AFM1 and interlayer FM coupling (i.e. AFM1-FM ground state). In contrast, the pristine bilayer stabilizes in the AFM1-AFM configuration. Specifically, for the Li-intercalated $V_2Se_2O$ bilayer, AFM1-AFM and FM-AFM states are 29.69 and 418.95 meV higher than the AFM1-FM ground state. For V-intercalation, these energy differences are 206.91 meV and 707.68 meV. Meanwhile, the pristine $V_2Se_2O$ bilayer shows an AFM1-AFM state that is 2.34 meV lower than the AFM1-FM state, and FM-AFM state is 917.41 meV higher. Since the comparison has already been demonstrated in Figure 3a, the energy values of other magnetic states will not be detailed here for brevity. Figure 3c outlines magnetic exchange interactions. In-plane V-V interactions are directly linked and indirectly O/Se-mediated. Out-of-plane couplings are Se-mediated in the pristine $V_2Se_2O$ bilayer, and additionally involve Li/V atoms in intercalated structures. These competing exchange interactions determine the magnetic ground state.

Atomic magnetic moments in the ground states are unveiled in Figure 3d, with provided atomic numbering (Figure 1a). The Li-intercalated $V_2Se_2O$ bilayer exhibits a small net magnetic moment of 0.384 $\mu_B$ per V atom pair, while the V-intercalated one has 0.662 $\mu_B$. The pristine bilayer remains fully compensated. However, the intralayer FiM orientation with small net magnetic moments in these intercalated bilayers remains analogous to the intralayer AFM1 configurations. To ensure consistency and facilitate comparison with the pristine bilayer, which also displays intralayer AFM1 characteristics, we have labeled these FiM configurations as AFM1 in Figures and Tables. The V atoms in pristine $V_2Se_2O$ bilayer have magnetic moments of approximately 2 $\mu_B$. After Li/V intercalation, the local magnetic moments of these V atoms change little, so these native V atoms in pristine and intercalated $V_2Se_2O$ bilayers are the $V^{3+}$ ($d^2$) state. Notably, intercalated Li and V manifest magnetic moments of near-zero and -0.903 $\mu_B$, respectively, so the intercalated V atom in $V_2Se_2O$ bilayer is the $V^{4+}$ ($d^1$) state. Intercalation induces charge transfer, modulates the interlayer distance and intralayer lattice, and further modifies the magnetic exchange interactions (Figure 3e).

Compared to AFM/AM materials, FiM materials endow several critical advantages for spintronics: (1) a small net magnetic moment ensures magnetization, weak stray field, and higher magnetic flux.[14] This makes them advantageous for scenarios where interaction with magnetic fields is essential. (2) Low coercivity[14,15] facilitates rapid and easy magnetization and demagnetization. This is particularly beneficial for signal reading and writing processes. (3) Momentum- and energy-dependent spin splitting may yield high spin polarization. The electronic structures and spin transport properties of FiM materials will be discussed in more detail later on. Furthermore, interlayer FM couplings in intercalated structures preserves and enhances electronic behavior by avoiding cancellation between constituent monolayers, unlike the interlayer AFM couplings found in pristine $V_2Se_2O$ bilayer.

### 2.2.2. Magnetic Anisotropy

Having established the ground-state magnetic configurations, we now analyze magnetic anisotropy, which stabilizes long-range magnetic order in 2D systems. The representative *x*-, *y*-, and *z*-orientations of the easy axes are compared (Figure 3b). The pristine $V_2Se_2O$ bilayer favors in-plane magnetization, with nearly isotropic *x*- and *y*-energies (-0.104 meV $f.u.^{-1}$). Intercalating preserves the in-plane preference but induces planar anisotropy: the ground state is tuned to *y* and *x* for Li- and V-intercalated $V_2Se_2O$ bilayers, respectively. For Li- (V-)intercalation, the energies of the easy magnetization axes along the *x*- and *y*-directions relative to the *z*-direction are -0.225 (-0.369) and -0.302 (0.046) meV $f.u.^{-1}$, respectively. Enhanced magnetic anisotropy improves resistance to thermal perturbations in a realistic RT environment. Notably, this in-plane transition from isotropy to uniaxial anisotropy supports fixed magnetic moments, and is critical for non-volatile memory.

The change in magnetic anisotropy can be further interpreted by examining the atomic and orbital contributions. Figure 4a-c displays the atomic contributions to magnetic anisotropy. V atoms dominate in all structures. Se atoms contribute notably in intercalated systems, whereas O atoms remain negligible. For V-intercalation, the added V atom plays a stronger dominant role than the native ones. In pristine $V_2Se_2O$ bilayer, anisotropy originates solely from V atoms while both O and Se atoms exhibit minimal endowments. In Li-intercalated bilayer, corner V2 and V4 dominate energy stabilization along the *y*-axis, while central V1 and V3 contribute more along the *x*-direction. Both V and Se atoms dedicate more along the *y*-direction than along the *x*-direction, making the *y*-axis the one with the lowest energy. For V-intercalated structures, the intercalated V atom shows the largest negative value, while the Se atoms possess positive values, resulting in the energy along *y*-direction being positive relative to the *z*-direction. The pristine $V_2Se_2O$ bilayer, although energetically isotropic in-plane, displays distinct atomic contributions: the central V1 and V3 (corner V2 and V4) atoms are dominant along the *x*- (*y*-) direction.

Employing the second-order perturbation theory,[73] the atomic and orbital origins are further clarified in Supporting Information. The V *d*-*d* and Se *p*-*p* interactions contributing to magnetic anisotropy are unveiled in Figure 4d-f, S3a,b and S4a,b.

For the Li-intercalated $V_2Se_2O$ bilayer (Figure S3a,b), the most stable configuration occurs when the magnetic moments align along the *y*-direction. This stability primarily arises from the hybridization of ($d_{xy}$, $d_{x^2-y^2}$), ($d_{xz}$, $d_{x^2-y^2}$) and ($d_{z^2}$, $d_{xz}$) in corner V2 and V4 atoms, which endow major contributions. In contrast, the hybridized ($d_{xy}$, $d_{yz}$) at these sites offers only a minor positive endowment, indicating a slight resistance to the *y*-direction easy axis. The hybridization contributions of ($d_{xz}$, $d_{x^2-y^2}$) and ($d_{z^2}$, $d_{xz}$) are positive, causing the central V1 and V3 to devote less than the corner V2 and V4. Additionally, the ($p_y$, $p_x$) and ($p_z$, $p_x$) interactions in Se atoms play a dominant role in stabilizing the magnetic anisotropy.

For the V-intercalated $V_2Se_2O$ bilayer (Figure 4d-f), the lowest-energy easy axis shifts to the *x*-direction, primarily owing to the contributions from the intercalated V rather than the native sites. The hybridization of ($d_{xy}$, $d_{x^2-y^2}$) and ($d_{z^2}$, $d_{xz}$) in the intercalated V possesses relatively larger endowment. Meanwhile, Se1 and Se4 exhibit positive valuations, mainly caused by their hybridized ($p_z$, $p_x$), which destabilize the *y*-direction alignment and make it the highest-energy, most unstable state.

In the pristine $V_2Se_2O$ bilayer (Figure S4a,b), the in-plane magnetic anisotropy is energetically isotropic, and it is more stable than the out-of-plane direction. However, the orbital hybridizations due to the SOC effect differ between these directions. When the easy axis is along the *x*-direction, dominant contributions arise from the hybridizations of ($d_{xy}$, $d_{x^2-y^2}$) and ($d_{yz}$, $d_{z^2}$) in the central V1 and V3, and ($d_{xy}$, $d_{x^2-y^2}$) in the corner V2 and V4. At the same time, hybridizations of ($d_{xy}$, $d_{xz}$) in the central V1 and V3 and ($d_{yz}$, $d_{z^2}$) in the corner V2 and V4 oppose alignment along the *x*-direction. When the easy axis is oriented along the *y*-direction, the main orbital donations of central V1/V3 and corner V2/V4 are effectively reversed. The orbital hybridization patterns for non-ground-state contributions follow similar trends and will not be discussed in detail here.

### 2.2.3. Magnetic Critical Temperature

For practical applications in spintronics, achieving a RT or above-RT magnetic critical temperature is highly desirable to ensure stable magnetic ordering. To investigate the temperature effect on the intra-FiM (AM) and inter-FM (AFM) configurations of $V_2Se_2O$ bilayers with (without) intercalation, we utilize Metropolis Monte Carlo (MC) simulations to estimate their magnetic critical temperatures. These critical temperatures are identified by the maximum slope in the magnetic moment versus temperature curve, and the peak of specific heat capacity.

The magnetic exchange interactions are modeled using Heisenberg spin Hamiltonian, which can be attained as,

$$H = -\sum_{<i,j>} J_1 \vec{S}_i \vec{S}_j - \sum_{<k,l>} J_{2x} \vec{S}_k \vec{S}_l - \sum_{<m,n>} J_{2y} \vec{S}_m \vec{S}_n - \sum_{<o,p>} J_{11} \vec{S}_o \vec{S}_p - \sum_{<q,r>} J_{int} \vec{S}_q \vec{S}_r - A\sum_i (\vec{S}_i^z)^2 \quad (1)$$

where the exchange interaction parameters $J_1$, $J_{2x}$, $J_{2y}$, $J_{11}$ and $J_{int}$ are defined in Figure 3c. The parameter *A* captures magnetic anisotropy, and $\vec{S}_i$ and $\vec{S}_i^z$ are the spin operator and its spin component parallel to the *z*-direction, respectively. Although the absolute values of individual spins differ, the net spin for a V-V pair is small. For consistency across systems, spin operators are normalized as $S_N = 1$ to solve for the exchange interaction parameters from the total energy expressions of magnetic states (Table S1). Since some magnetic states share the same expressions, our MC simulations focus on ground states and low-energy states close to the ground states. Validation for our MC approach's reliability can be seen in Supporting Information.

With normalized parameters of magnetic exchange interactions (Table 2), the critical temperatures are simulated in Figure 3f-h. The Li-intercalated, V-intercalated, and pristine $V_2Se_2O$ bilayers possess robust above-RT magnetic critical temperatures of 358, 773, and 960 K, respectively. These observed robust above-RT critical temperatures with and without intercalations are higher than those of most known 2D magnetic materials,[36,74] uncovering the potential for practical magnetic applications of $V_2Se_2O$-based systems.

The variation in magnetic critical temperatures can be attributed to the modulated exchange interactions. Compared with the pristine $V_2Se_2O$ bilayer, both Li- and V-intercalations lead to the emergence of interlayer FM couplings with positive $J_{int}$ values, replacing the AFM interactions found in the unmodified system. Briefly, in the Li-intercalated system, in-plane magnetic exchange interactions $J_1$ and $J_{11}$ weaken, and the $J_{2x}$ changes from negative to positive, turning AFM to FM interaction. The overall effect results in a lower magnetic critical temperature than the pristine bilayer, despite a smaller interaction $J_{2y}$. In the V-intercalated system, $J_1$ and $J_{2x}$ interactions are enhanced (stronger FiM interaction), and $J_{2y}$ reverses to FiM from FM. However, both $J_{11}$ and $J_{int}$ switch from AFM to FM, causing a moderate reduction in magnetic critical temperature compared with the non-intercalated bilayer. Magnetic exchange interactions $J_{2x}$, $J_{2y}$, and $J_{11}$ show differences, and similar signs have also been reported.[52] Owing to the calculation methods for magnetic exchange interactions detailed in Supporting Information, their signs are determined not only by the magnetic ground-state configuration but also by other magnetic states considered. These differences come from the used energies of magnetic states (Figure 3a) for calculating magnetic exchange interactions. Intercalating atoms have an impact on both intralayer and interlayer magnetic exchange interactions, resulting in different energy differences between various magnetic states. The drastic change between pristine $V_2Se_2O$ (958 K) and Li-intercalated sample (395 K) mainly stems from the significant change in $J_{2y}$, which is 155.909 meV f.u.$^{-1}$ for pristine $V_2Se_2O$ while 38.771 meV f.u.$^{-1}$ for Li-intercalated sample (Table 2). This variation in $J_{2y}$ arises from the altered energy difference between the AFM5-FM and AFM6-FM states, which is 0.630 eV in pristine $V_2Se_2O$ and 0.086 eV in Li-intercalated sample (Figure 3a). The V-V distances corresponding to these considered magnetic exchange interactions are listed in Table 1. Distances $d_1$, $d_{11}$ and $d_{2x}$ all increase upon intercalation, particularly in V-intercalation. Conversely, $d_{2y}$ values decrease in both intercalated cases, with a more also noticeable drop in the V-intercalated $V_2Se_2O$ bilayer. Additionally, the interlayer Se-Se distances decrease in both Li- and V-intercalation, the interlayer V-V distance ($d_{int}$) decreases with Li-intercalation, but increase with V-intercalation, indicating distinct vertical structural modifications induced by different intercalants.

### 2.3. Ferroelastic Switching

FC materials undergo lattice reorientation under stress, driven by a spontaneous strain that acts as an order parameter. This strain induces crystallographic symmetry changes, determining the material's macroscopic state. The FC switching process is characterized to reversible crystal distortion.[75] FC materials, similar to magnetic materials, display domain structures and hysteresis upon stress removal. Their ability to intertwine mechanical motion with electrical energy enables reversible mechanical actuation, making them valuable for signal conversion, sensitive sensing, and electromechanical applications.[75,76]

Ferroelasticity requires that a material has two or more stable orientation states, which can be switched by external stress. The orthogonal in-plane lattice constants $a$ and $b$ should be unequal to allow stress-induced switching between orientations—a necessary condition for FC switching in 2D materials, as verified by previous experimental and theoretical studies.[20,77,78] Li- and V-intercalated $V_2Se_2O$ bilayers possess two degenerate ground states with orthogonal lattice orientations, related by a 90° rotation, which are switchable under external uniaxial strain. Rooted in the acoustic mode softening theory,[79] ferroelasticity originates from the instability of a highly symmetric phase. Intercalation disrupts the chemical environment, modifies the in-plane lattice constants, leading to ferroelasticity. The initial and final FC states are denoted as $F_1$ ($a > b$) and $F_2$ ($a < b$), respectively. Transition between these states occurs via a metastable, para-elastic state $P$ ($a = b$). However, for the pristine $V_2Se_2O$ bilayer, $a$ is equal to $b$, preventing stress-induced switching and thus the emergence of FC behavior.

The evaluation of ferroelasticity centers on two key criteria: the switching barrier and the reversible ferroelastic strain. The energy barrier should be above ~0.03 eV for robustness against RT thermal fluctuations and below ~0.6 eV for feasible switching under ambient conditions.[78] The energy barriers for FC switching in Li- and V-intercalated systems are 0.210 and 0.202 eV (Figure 5a), respectively. These barriers in our study are comparable to phosphorene (0.200 eV)[75] and GaSb (0.225 eV),[78] smaller than $BP_5$ (0.320 eV)[80] and VOCl (0.256 eV).[81] The FC switching pathways (Figure 5b,c) confirm only the intercalated systems exhibit FC switching, whereas the pristine bilayer does not. The corresponding structures of initial $F_1$, metastable $P$ and final $F_2$ states clearly reveal the nature of the phase transition.

The FC signal intensity, denoting the strength of the FC response, can be evaluated by the reversible FC strain $\varepsilon$ (Table 1), defined as $\varepsilon = (|a| / |b| - 1) \times 100$ %. For Li- and V-intercalated $V_2Se_2O$ bilayers, $\varepsilon$ is 0.616 % and 1.990 %, respectively. These strains are well within experimental feasibility—compared to the 6 % strain achievable in graphene[82] and 0.2 %–2 % strain range accurately implemented in $MoS_2$.[83] Our values fall in the typical FC range (0.5 % - 3 %),[75,76] and are comparable to $ReS_2$ (1 %),[84] but lower than those in common FC materials like ZnS (12.8 %)[85], phosphorene (37.9 %),[75] GaSb (67.0 %),[78] $BP_5$ (41.4 %),[80] and VOCl (15.1 %).[81] A small reversible strain, implying a weaker FC signal, presents advantages such as enhanced sensitivity to minor stimuli, lower switching thresholds, higher conversion efficiency, lower power consumption, and reduced material fatigue. These properties are crucial for ensuring a long service lifetime and seamless integration into multifunctional systems.

In light of above results, the ferroic evolution of $V_2Se_2O$ driven by intercalation can be drawn as Figure 5d. Band structure analyses (Figure 6a-f) confirm the AM-type spin splitting in the pristine $V_2Se_2O$ bilayer, which is absent in the Li- and V-intercalated bilayers. The pristine $V_2Se_2O$ bilayer is single-ferroic, exhibiting intralayer AM and interlayer AFM couplings in its ground state. After Li- and V-intercalations, the system becomes multiferroic, showing intralayer FiM and interlayer FM couplings, along with ferroelasticity. These findings furnish a platform for exploring magnetoelastic coupling, expanding the theoretical boundaries of the AM family, and promoting the exploitation of potential multifunction by tuning altermagnets. Electrochemistry-intercalation provides a reversible and efficient route to control material properties, suggesting potential for electrical and chemical energy conversion in AM family. Self-intercalation demonstrates excellent compatibility and feasibility for experimental realization. Overall, intercalation emerges as a powerful solution for expanding the functionality of AM materials, developing multiferroic features and paving the avenue of compact integrated devices.

### 2.4. Enhanced Spin Splitting and Half Metal

High spin polarization is crucial for efficient spin filtering, spin transport and magnetic data storage. However, the spin splitting of altermagnets is momentum-dependent and energy-independent, typically restricted to a few specific high-symmetry points in the Brillouin zone. Consequently, practical spintronics generally require special orientations or engineered interfaces to harness spin-dependent behavior. In other cases, AM system exhibits conventional AFM-type spin degeneracy with zero spin polarization, severely limiting their pragmatic applicability. This raises a critical question: can intercalation modify the spin-resolved electronic structure to enhance spin splitting and achieve high spin polarization?

To answer this, the band structures of Li-intercalated, V-intercalated and pristine $V_2Se_2O$ bilayers are examined along different crystallographic pathways (Figure 6a-f), with corresponding high-symmetry points displayed in Figure 6g. The pristine $V_2Se_2O$ bilayer, with intralayer AM and interlayer AFM couplings, is an AM semiconductor, and its band gap of 0.68 eV is similar to that of its monolayer counterpart. Intercalation induces an electronic phase transition. Specifically, the Li-intercalated $V_2Se_2O$ bilayer with intralayer FiM and interlayer FM state becomes metallic, with bands in both spin channels clearly crossing the Fermi level. Similar electronic and magnetic transitions caused by intercalation have been reported in some experimental systems,[50,55,67,68] supporting our results. Spin splitting is significantly enhanced at high-symmetry points Γ, M, X, and Y (Figure 6a). In contrast, the pristine bilayer exhibits splitting only near the X and Y points (Figure 6c). Its energy projection yields identical electronic states for both spin channels, resulting in negligible spin filtering.

While Li-intercalation enhances spin splitting, the spin polarization remains low owing to minimal differences in the band numbers in the two spin channels across the Fermi level. Fortunately, V-intercalation induces a HM feature: the spin-up channel becomes conductive, while the spin-down channel opens a band gap of 0.97 eV (Figure 6b), allowing for a perfect (100 %) spin polarization. Thus, the V-intercalated $V_2Se_2O$ bilayer is a promising candidate for spintronic nano-devices that require high-efficiency spin filtering.

Out-of-plane band dispersions along the Γ—Z path (Figure 6d-f) further illustrate spin-resolved electronic behaviors. For the Li-intercalated system, both spin channels have states near the Fermi level ($-0.04$ eV $< E <$ 0.26 eV), which generates inferior spin polarization in equilibrium. However, shifting the Fermi level enhances energy-dependent spin splitting, although downward shifts yield minimal improvement (Figure 6d). The V-intercalated $V_2Se_2O$ bilayer remains its HM character, with abundant spin-up states close to the Fermi level (Figure 6e). For comparison, the pristine system shows spin-degenerate throughout (Figure 6f), even under Fermi level adjustments. These results uncover the significant impact of intercalation on the electronic structure and spin polarization of $V_2Se_2O$ bilayers.

The atomic contributions to the electronic structures are analyzed via the atom- and spin-resolved band structures (Figure S5a-p and S6a-f). In all cases, V atoms dominate both valence and conduction bands around the Fermi level, while Se atoms primarily influence the valence band. In the V-intercalated $V_2Se_2O$ bilayer, the intercalated V atom plays a key role in forming the HM feature, whereas the intercalated Li atom contributes negligibly to the Li-intercalated $V_2Se_2O$ bilayer's electronic structure. Further V-*d*- and Se-*p*-orbital-resolved band structures (Figure S7-S11) reveals the following aspects: For Li-intercalated $V_2Se_2O$ bilayer, conduction is governed by the $d_{xz}$ orbital in the spin-up channel and the $d_{yz}$ orbital in the spin-down channel of V atoms. Near the Fermi level, V-$d_{xy}$ and Se-$p_y$, $p_z$ and $p_x$ orbitals in both spin channels noticeably contribute to the valence bands, while V-$d_{x^2-y^2}$ orbitals in both spin channels, V-$d_{yz}$ orbital in the spin-up channel and V-$d_{xz}$ orbital in the spin-down channel in the conduction bands are also noticeable. In the V-intercalated $V_2Se_2O$ bilayer, the conductive spin-up channel is attributed by $d_{xz}$ orbital of the native V and $d_{x^2-y^2}$ orbital of the intercalated V, while the semiconductive spin-down channel involves V-$d_{yz}$ orbital in the valence bands, and V-$d_{xy}$ and Se-$p_y$ orbitals in the conduction bands. For the pristine bilayer, the valence band maximum stems mainly from V-$d_{yz}$ and $d_{xz}$ orbitals, while its conduction band minimum is primarily devoted by V-$d_{xy}$ and Se-$p_y$, $p_z$ and $p_x$ orbitals.

The transition from semiconductor to metal is attributed to electron transfer through intercalation. Band structures contributed by each V atom (Figure S12-S17) and intercalated V (Figure S7) display the alternating spin orientations: V1 and V3 atoms are spin-down; V2 and V4 atoms are spin-up. Each V atom in the $V_2Se_2O$ bilayer has a magnetic moment close to 2 $\mu_B$. The occupied $d_{xy}$ state possesses two almost overlapping bands in the valence band near the Fermi level, while the unoccupied states are visible in $d_{yz}$, $d_{xz}$ and $d_{x^2-y^2}$ orbitals. After intercalation, V atoms gain a small number of electrons, verified by atomic magnetic moments in Figure 3d. These allow partially occupied states to appear, there are bands across the Fermi level in intercalated $V_2Se_2O$ bilayers. For the Li-intercalated $V_2Se_2O$ bilayer, the $d_{yz}$ ($d_{xz}$) orbital in the spin-down (up) channel possesses the partially occupied states, meaning metallicity in both spin channels. In the V-intercalated $V_2Se_2O$ bilayer, the $d_{xz}$ orbital in the spin-up channel is partially occupied in the native V atoms, and peculiarly, the $d_{x^2-y^2}$ orbital of the intercalated V atom is also partially occupied, but the spin-down channel doesn't show occupation, resulting in half-metallicity.

A schematic of these electronic transitions via intercalation is illustrated in Figure 6h. The pristine $V_2Se_2O$ bilayer is an AM semiconductor. After intercalation, it becomes metallic. Metals typically offer higher conductivity and tunability than semiconductors, making them advantageous for practical spintronic devices, especially for modulating the direction of spin currents through an electric field.[50] These results demonstrate the advantages of Li- and V-intercalated $V_2Se_2O$ bilayers over the pristine system. Considering the spin behavior, the pristine bilayer possesses momentum-dependent and energy-independent spin splitting, which is inconveniently exploited. The enhanced spin splitting and HM traits can be secured with Li- and V-intercalations, respectively. V-intercalated $V_2Se_2O$ bilayer can yield complete spin filtering, and that of Li-intercalation is poor but will be elevated when the Fermi level is shifted. It's uncovered that Li- and V-intercalated AM $V_2Se_2O$ bilayers can be employed in miniaturized spintronics to accomplish high spin filtering efficiency and eminent magnetic resistance.

## 2.5. Anomalous Hall Effect

The AHE, which arises from intrinsic material magnetization and SOC, is important for low-power detection and sensing technologies. SOC is particularly valuable in maintaining quantum coherence, driving AHE-related phenomena relevant to both magnetic storage and quantum computing.

The AHE behavior in pristine, Li-intercalated, and V-intercalated $V_2Se_2O$ bilayers is systematically examined. The anomalous Hall conductivity (AHC)[86] can be written as,

$$\sigma_{xy} = \frac{e^2}{\hbar}\left(\frac{1}{2\pi}\right)^3 \int_{BZ} \Omega_z(k) d^3k \quad (2)$$

where $e$ is the electron charge, $\hbar = h/2\pi$ stands for the reduced Planck's constant, $k$ is the crystal momentum, and $\Omega_z(k)$ is the Berry curvature, discussed in more detail below. The calculated energy-dependent AHC (Figure 7a) remains modest near the Fermi level in all systems. However, these AHC values could be enhanced through approaches such as interfacial engineering,[87] external magnetic and electric fields,[88,89] and strain engineering,[90] which are in favor of practical utilization and merit further investigation. The sign and magnitude of AHC are determined by magnetic moment orientation, charge carrier type, and their mutual interactions. Reversing the spin orientation or charge polarity flips sign of the transverse Hall conductivity. Electrons and holes contribute oppositely, and the magnetic moment direction also modulates carrier dynamics via exchange interactions. Moreover, metallic systems offer advantages such as higher carrier concentrations, conductivity, and thermal stability, which are beneficial for AHE. However, their typically lower Hall voltage may challenge detection sensitivity, representing a trade-off that warrants future investigation.

To explain the origin of AHE in $V_2Se_2O$ bilayers with and without intercalations, we use the Kudo formula[91] for the *k*-resolved Berry curvature $\Omega_z(k)$, which can be attained as $\Omega_z(k) = \sum_n f_n \Omega_n(k)$, in which $f_n$ is the Fermi-Dirac distribution. The Berry curvature, integrated over all occupied states, is central to understanding the AHE, which can be written as,

$$\Omega_n(k) = -2\mathrm{Im}\sum_{m\neq n} \frac{\hbar^2 \langle \Psi_{nk} | v_x | \Psi_{mk} \rangle \langle \Psi_{mk} | v_y | \Psi_{nk} \rangle}{\left(E_m(k) - E_n(k)\right)^2} \tag{3}$$

where the $E_m$ stands for the eigenvalue of the Bloch function $|\Psi_{mk}\rangle$, and $v_x$, $v_y$ are the velocity operators along the *x*- and *y*-directions, respectively.

The momentum-dependent Berry curvature distributions for each system are illustrated in Figure 7b-d. Distinct peaks are observed around high-symmetry points: sharp peaks near M and X for Li-intercalated, localized peaks near M for V-intercalated, and strong contributions at M and Γ for non-intercalated $V_2Se_2O$ bilayers. These localized Berry curvature peaks confirm the intrinsic AHE signatures in all three systems, in accordance with the trends of AHC (Figure 7a). The distribution of Berry curvature in momentum space governs the transverse motion of charge carriers, hence directly influencing the AHE. Intercalation modifies the electronic structure and crystal symmetry, thereby redistributing the Berry curvature in momentum space and consequently influencing the AHC. Similar transport properties modified by intercalation have been experimentally reported.[54,55,67,92] This tunability establishes intercalation as an effective tool of tailoring the AHE, particularly for low-consumption detection and sensing applications. Future research could explore how different intercalation strategies or external stimuli further optimize the AHE, potentially improving the anomalous transport technologies.

The modified momentum- and spin-dependent states around the Fermi level for each bilayer configuration (Figure 7e) provide further insights. The corresponding AHE device models incorporating pristine, Li-intercalated and V-intercalated $V_2Se_2O$ bilayers are illustrated in Figure 7f-h. In the pristine and Li-intercalated bilayers, contributions from carriers at the X and Y points partially cancel. However, a net Berry curvature remains because the negative region is larger than the positive region (Figure 7b and 7d), which leads to a relatively large total AHC. Interestingly, these systems exhibit a momentum differentiated spin behavior. The spin-down (up) channel dominate at the X (Y) point in the pristine bilayer, regardless of carrier type. Spin-up (down) electrons at X (Y) points can accumulate on one (the other) side, with holes behaving oppositely at these symmetry points, giving rise to charge-spin-momentum locking. This phenomenon couples spin, momentum, and charge, potentially enabling enhanced control of carrier trajectories via external fields. In contrast, the V-intercalated bilayer exhibits constructive contributions to the AHE from both X and Y points, with a more uniform distribution in momentum space. But the sum of Berry curvature is close to 0 (Figure 7c), resulting in a very small total AHC. These AHE characters make the pristine and intercalated $V_2Se_2O$ attractive for Hall components.

### 2.6. Multistate Spin Transport

#### 2.6.1. Equilibrium Transport Performance

Spintronics offers remarkable pathways for information storage, transmission, and processing. Two cornerstone effects – GMR and TMR – enable distinct functionalities. GMR originates from the relativistic effect of spin-dependent electron scattering, typically employing metallic non-magnetic intermediate layers to deliver high signal outputs. In contrast, TMR is governed by quantum tunneling through insulating or semiconducting barriers, excelling in sensitivity and stability. Both effects have driven advances in magnetic random-access memory, read-write heads, and magnetic sensors.

Given the enhanced spin splitting and HM characteristics of intercalated $V_2Se_2O$ bilayers, we propose corresponding spintronic devices (Figure 8a-c), with the pristine bilayer serving as a baseline for comparison. These devices incorporate intermediate layers of Au and $SrZrO_3$ and are fully relaxed. Gold is selected as an electrode due to its excellent electrical conductivity and widespread use as a substrate for experimental growth of 2D magnetic materials.[74,93] The electrodes are designed solely for electrical conduction, with no additional functional effects; thus, other non-magnetic conductors may also be suitable provided that lattice matching is satisfied. The intermediate layers consist of conductive Au and semiconducting $SrZrO_3$ perovskite, chosen to potentially realize GMR and TMR, respectively. Crystal configurations and band structures of Au and $SrZrO_3$ are demonstrated in Figure S18. Bulk and monolayer Au exhibit metallic behavior, while bulk $SrZrO_3$ has a band gap of 3.31 eV, consistent with previous reports;[94,95] its monolayer shows a moderate band gap of 1.19 eV. A vacuum layer is included as a reference to evaluate the influence of the intermediate layer. The lattice constants of Au and $SrZrO_3$ are 4.155 and 4.174 Å, respectively, while those of with Li-, V-, and non-intercalated $V_2Se_2O$ bilayers are provided in Table 1. Lattice mismatch is defined as $\Delta = |a_2 - a_1|/a_1 \times 100\%$, where $a_1$ and $a_2$ are the respective lattice constants. The calculated mismatch values for the devices (Figure 8a-c) are 2.33 %, 2.78 %, 2.33 %, 1.32 %, 1.77 %, 1.32 %, 2.77 %, 3.21 %, and 2.77 %, all of which are experimentally feasible,[96,97] confirming structural compatibility. Device performance is evaluated under both parallel (P) and antiparallel (AP) magnetization alignments, referring to identical and opposite magnetic configurations of the left and right $V_2Se_2O$-based bilayers, respectively.

In equilibrium, the spin filtering efficiency $\eta$ is defined as $\eta = (T_{\uparrow} - T_{\downarrow}) / (T_{\uparrow} + T_{\downarrow}) \times 100\%$, where $T_{\uparrow}$ and $T_{\downarrow}$ are the transmission coefficients for the spin-up and spin-down channels, respectively. Positive and negative values of $\eta$ correspond to filtering spin-down and spin-up channels, respectively. The magnetoresistance (MR) ratio is given by $\mathrm{MR} = (T_{\mathrm{P}} - T_{\mathrm{AP}}) / T_{\mathrm{AP}} \times 100\%$, where $T_{\mathrm{P}}$ and $T_{\mathrm{AP}}$ are the total transmission coefficients under P and AP alignments, respectively. GMR occurs with metallic spacers, while TMR arises with semiconducting or insulating layers. Detailed formalisms are presented in Supporting Information. These metrics quantitatively characterize spin transport performance.

Spin-resolved transmission coefficients as a function of energy are shown in Figure 9a,b and S19a,b. The $\vec{k}_{//}$- and spin-resolved transmission spectra in the 2D Brillouin zone are illustrated in Figure 9c-z and S19c-n. For notational simplicity, the proposed MTJs are denoted as Li/V/Bil-Au/SZO/Va, representing Li-, V-, and non-intercalated $V_2Se_2O$ bilayers with Au, $SrZrO_3$, and vacuum intermediate layers, respectively. Table 3 details $T_{\uparrow}$ and $T_{\downarrow}$, $\eta$, and MR ratios for all devices in equilibrium. To elucidate transport mechanisms, we also compute the spin-resolved local density of states (LDOS) (Figure 10a-x and S20a-l), and the spin- and layer-resolved projected device density of states (PDDOS) (Figure 11a-f, S21a-f, and S22a-f), with the left Au electrodes serving as the source of contribution. Notably, interface effects lead to unequal electronic state projections in the two spin channels for non-magnetic Au and $SrZrO_3$, as well as for the AM $V_2Se_2O$ bilayer.

For Au/Li@$V_2Se_2O$/Au/Li@$V_2Se_2O$/Au, transmission is concentrated near the X and Y points and away from these points, with minimal contribution around Γ, M, or other points (Figure 9c-f). In contrast, transmission is primarily near the Γ point in devices with vacuum or $SrZrO_3$ (Figure 9k-n and 9g-j). Across all configurations, P and AP alignments yield similar transmission at the Fermi level in both spin channels, resulting in low spin filtering efficiency and MR ratios. The equilibrium electronic states show negligible differences between spin channels or magnetization configurations (Figure 10a-h and S21a-d), consistent with the limited performance. The Li-intercalated bilayer exhibits low spin polarization despite enhanced spin splitting (Figure 6a,d), indicating inferior spin filtering. Notably, a vacuum intermediate layer enhances spin polarization: the P configuration displays dominant spin-down transmission, yielding $\eta$ = 82 %. However, AP configuration exhibits minimal spin transport and low spin filtering efficiency. A high TMR of 677 % is achieved with a vacuum spacer in the Li-intercalated system, attributable to a peak in the spin-down LDOS of the P state near the Fermi level in the extended region of the right Au electrode, while the spin-up channel in the P state and both channels in AP alignment lack available states (Figure S21e,f). These results indicate that interface effects between the Li-intercalated $V_2Se_2O$ bilayer and the extended region of Au electrodes, combined with a vacuum spacer, significantly enhance spin polarization. Nevertheless, the generally low spin polarization suggests that further optimization—such as interface engineering and alternative intercalation strategies—is needed to improve performance.

For devices with V-intercalated $V_2Se_2O$ bilayers, transmission is mainly concentrated away from X and Y points along off-high symmetry directions (Figure 9o-z), and in few cases (like the spin-up channel of device using a vacuum intermediate layer in Figure 9w), it is concentrated close to M points. Regardless of the intermediate layer, the P state achieves near-perfect spin filtering efficiency ($\eta$ = 96 % - 98 %), while the AP state performs less effectively. High MR ratios of 877 % (Au), 716 % ($SrZrO_3$), and 1194 % (vacuum) are obtained due to superior spin-up transmission in the P state compared to the spin-down channel in the P state and both channels in the AP state. The HM feature of V-intercalated bilayer (Figure 6b,e) underpins this high performance. In the P state, more electronic states in the spin-up channel through the extended region of the right electrode than in the spin-down channel due to the half-metallicity and matched conductivity in the spin-up channel for left and right V-intercalated $V_2Se_2O$ (Figure 10m,n,q,r,u,v and 11a,c,e), and significantly more than in the AP state for both spin channels due to the mismatched conductivity between left and right V-intercalated $V_2Se_2O$ (Figure 10o,p,s,t,w,x and 11b,d,f). This leads to high $\eta$ in the P state and high MR ratios overall. In the AP state, little difference in electronic states between spin channels results in less prominent spin filtering. The electronic structure of V-intercalated system remains robust across different intermediate layers, highlighting its reliability as a platform for miniaturized spintronic devices.

Across non-intercalated $V_2Se_2O$ bilayers, Au spacers yield transmission around the X and Y points (Figure S19c-f), while $SrZrO_3$ and vacuum concentrate transmission near the Γ point (Figure S19g-n). For all intermediate layers, transport coefficients show minimal variation across magnetization configurations and spin channels, resulting in low $\eta$ (≤ 23 %) and negligible MR (≤ 11 %). A negative TMR occurs with $SrZrO_3$ due to slightly higher transmission in AP state than in P state. The pristine $V_2Se_2O$ bilayer's AM nature leads to momentum-dependent but energy-independent spin splitting and out-of-plane AFM-type spin degeneracy (Figure 6c,f), making high spin filtering and MR challenging. Similar electronic states in both spin channels and magnetization configurations (Figure S20a-l and S22a-f) result in low performance regardless of the intermediate material.

Figure 8d summarizes the enhanced spin transport performance of $V_2Se_2O$ bilayers achieved via intercalation engineering. The non-intercalated system displays minimal GMR, TMR and $\eta$ due to spin degeneracy. Li-intercalation moderately improves spin polarization, but significant TMR and filtering occur only with a vacuum spacer. In contrast, V-intercalated bilayers exhibit robust and high GMR/TMR and $\eta$ across all intermediate layers, attributable to their intrinsic HM character. These results highlight the V-intercalated $V_2Se_2O$ bilayer as a highly promising candidate for advanced spintronic devices.

### 2.6.2. Temperature Effect on Spin Current

The interaction between spin and thermal motion is central to thermal spintronics, which offers opportunities to harness waste heat and improve energy conversion efficiency. The calculated thermal spin transport properties for Li-, V-, and non-intercalated bilayers with Au, $SrZrO_3$, and vacuum spacers are presented in Figure 12a-d, S23a-d, and S24a-d. The corresponding formalisms are provided in Supporting Information. The left electrode temperature ($T_L$) is varied (250, 300, and 350 K), with temperature gradients $\Delta T$ = 20, 40, and 60 K. Since MR cannot be calculated under zero current when $\Delta T$ = 0, this point is omitted; equilibrium properties have been discussed in the previous section. Spin-resolved transmission coefficients are illustrated in Figure 9a,b. The Fermi-Dirac distribution difference $\Delta f$ (Figure 12e) increases with larger $T_L$ or $\Delta T$, indicating enhanced thermal current in a given spin channel. Table 4 summarizes performance at $T_L$ = 300 K and $\Delta T$ = 20 K as a representative case.

Devices with Li-intercalated $V_2Se_2O$ bilayers display variable $\eta$ in the P state: ≈ -75 % with Au near $T_L$ = 100 K; ≈ -90 % with $SrZrO_3$ around $T_L$ = 150—170 K; ≈ -100 % with vacuum across $T_L$ = 100—350 K or $\Delta T$ = -60—60 K. With Au and $SrZrO_3$, thermal MR ratios improve but remain below 60 %. Notably, with a vacuum spacer, thermal TMR reaches exceptionally high values: 11840 %, 11293 %, and 10454 % at $T_L$ = 100 K with $\Delta T$ = 60, 40, and 20 K, respectively. Although thermal TMR decreases with increasing $T_L$, the lowest TMR is still high with 2163 %, 2098 %, and 2037 % at $T_L$ = 350 K with $\Delta T$ = 60, 40, and 20 K, respectively. The Fermi-Dirac distribution and transmission spectrum explain these thermal current changes. For Li-intercalated systems with a vacuum spacer at $T_L$ = 350 K and $\Delta T$ = 60 K, the main $\Delta f$ spans a specific energy range from -0.18 to 0.18 eV (Figure 12e). In this range, the spin-down transmission coefficients in the P state are higher (Figure 9a,b), leading to higher thermal spin-down currents and high spin polarization. Additionally, transmission coefficients in both spin channels of the AP state are low (Figure 9a,b), leading to the high TMR.

Devices utilizing V-intercalated $V_2Se_2O$ bilayers consistently exhibit near-perfect spin filtering across all intermediate layers. Thermal MR ratios reach several hundred percent, with peak TMR values of 4355 % ($\Delta T$ = 20 K and $T_L$ = 140 K), 4337 % ($\Delta T$ = 40 K and $T_L$ = 150 K), and 4307 % ($\Delta T$ = 60 K and $T_L$ = 160 K) with a vacuum spacer. Even at higher temperatures, TMR remains around 760 % at $T_L$

≈ 300 K. Furthermore, as $T_L$ or $\Delta T$ increases, V-intercalated systems with Au and vacuum spacers exhibit the SSE (Figure 12a,b), where spin-up and spin-down currents carry opposite signs. The positive and negative signs indicate hole- and electron-dominated transport, respectively. The clarity of SSE depends on the comparability of the two spin currents—for instance, in V@$V_2Se_2O$-Au-P in the first sub-Figure of the first row in Figure 12a, the spin-up current is positive while the spin-down current is negative especially when $T_L$ is higher than 200 K, but the spin-up current is several orders of magnitude higher than spin-down current, so this SSE is not obvious. For V@$V_2Se_2O$-Au-AP in the second sub-Figure of the first row in Figure 12a, the spin-up and spin-down currents are approximately the same order of magnitude, and they show opposite signs, indicating the obvious SSE. For V@$V_2Se_2O$-Au-AP in the second sub-Figure of the first row in Figure 12b, the currents in the two spin channels are comparable and they show obviously opposite signs. For V@$V_2Se_2O$-SZO, there is no SSE. For V@$V_2Se_2O$-Va, the vacuum layer greatly hinders tunneling, so the currents are all small. For V@$V_2Se_2O$-Va-P in the first sub-Figure of the third row in Figure 12a, the currents in the two spin channels are roughly comparable and show opposite signs when 100 K < $T_L$ < 250 K. For V@$V_2Se_2O$-Va-P in the first sub-Figure of the third row in Figure 12b, the SSE occurs when $T_L$ = 250 K and $\Delta T$ > 0. This behavior arises due to the dominant transmission coefficients on opposite sides of the Fermi level within the effective $\Delta f$ interval in the two spin channels (Figure 9a,b and 12e).

Non-intercalated $V_2Se_2O$ bilayers devices with Au and $SrZrO_3$ show moderate $\eta$ ≈ 50–70 % with Au at $T_L$ = 100–130 K and 220–350 K (Figure S24). Nearly 100 % filtering occurs at $T_L$ = 170 K and $\Delta T$ = 20 K; $T_L$ = 180 K and $\Delta T$ = 40 K; and $T_L$ = 190 K and $\Delta T$ = 60 K, where the spin-up current nearly vanishes. Maximum GMR reaches -90 %, though MR is generally low. TMR with $SrZrO_3$ and vacuum remains below 10 %. Similar transmission in both spin channels results in poor MR and spin filtering. Moreover, at a certain $T_L$, the effective energy range of $\Delta f$ (Figure 12e) is larger for negative $\Delta T$ (-20, -40, and -60 K) than for positive $\Delta T$ (20, 40, and 60 K).

This allows more transmission coefficients to be included (Figure 9a,b), leading to larger thermal currents for negative $\Delta T$, a trend consistent across all materials.

Negative thermal MR occurs in Au/Li@$V_2Se_2O$/$SrZrO_3$/Li@$V_2Se_2O$/Au (Figure S23c) and in Au/$V_2Se_2O$/Au/$V_2Se_2O$/Au and Au/$V_2Se_2O$/$SrZrO_3$/$V_2Se_2O$/Au (Figure S24c,d), resulting from higher AP transmission within the effective $\Delta f$ window. Interestingly, some spin currents decrease with increasing $T_L$ or $\Delta T$, independent of sign convention—a phenomenon termed TNDR effect. Examples include: (1) For Li@$V_2Se_2O$-Va-P in the first sub-Figure of the third row in Figure S23, the spin-down current obviously decreases as $T_L$ rises from 130 K to 250 K. (2) For $V_2Se_2O$-Au-AP in the second sub-Figure of the first row in Figure S24a, the spin-up current decreases as $T_L$ rises from 120 K to 200 K. (3) For V@$V_2Se_2O$-SZO-AP in the second sub-Figure of the second row in Figure 12a, the spin-up current decreases with $T_L$ increasing from 130 K to 220 K. TNDR arises from that transmission coefficients decrease with changing energy (Figure 9a,b), and only the narrow $\Delta f$ window (Figure 12e) contributes to the thermal current. Beyond this range, even large coefficients do not contribute to thermal current. These changes in transmission coefficient can be further interpreted by electronic structure variations, where the amplitudes of spin-resolved electronic states differ as energy level changes. The TNDR effect may be applicable in RT microwaves and oscillators.

The thermal spin transport performance is summarized in Figure 12f. Non-intercalated bilayers exhibit limited spin filtering and MR. Li-intercalation enables high spin filtering and ultra-high TMR with a vacuum spacer. V-intercalated systems consistently achieve excellent performance across temperatures. SSE and TNDR effects are observed in intercalated systems, demonstrating their versatile and promising characteristics for RT thermal spintronics.

## 3. Conclusion

In conclusion, this work elucidates electrochemistry- and self-intercalation mechanisms in layered altermagnets, highlighting their potential for multifunctional integration (Figure 1e). The experiment-feasible layered AM $V_2Se_2O$ is selected as a representative system. Systematic investigations reveal that Li- and V-intercalated $V_2Se_2O$ bilayers exhibit above-RT intralayer FiM and interlayer FM coupling, as well as intrinsic ferroelasticity with a high signal intensity of ~1 %. Intercalation engineering effectively tailors the in-plane uniaxial magnetic anisotropy and transforms the electronic structure: Li-intercalation enhances spin splitting and induces metallicity, while V-intercalation imparts a HM feature. The AHE is observed in both pristine and intercalated systems. Spintronic and thermal spintronic devices based on intercalated bilayers demonstrate promising spin transport properties, achieving large GMR and TMR ratios, as well as high spin filtering efficiency (~100 %). Notably, a large thermal TMR (~12000 %) and a GMR (877 %) are achieved in the Li- and V-intercalated systems, respectively. These devices also exhibit the SSE and TNDR effect, highlighting their multifaceted thermal spintronic capabilities. Collectively, the interplay between intercalation engineering and the multifunctional properties of layered altermagnets suggests a promising strategy for developing advanced, miniaturized, and RT applications, such as low-consumption high-density non-volatile magnetic memory, sensors, and MEMS.

This work establishes a framework in which intercalation in a candidate altermagnet can induce modified electronic structures, multiferroic FiM-FC character, and anomalous and spin transport properties. This approach holds substantial promise and could be extended to other layered altermagnets. The insights from this study are expected to contribute to the expanding field of altermagnetism and motivate future work on multifunctional AM systems.

## 4. Experimental Section

The first-principles calculations were carried out by the density functional theory (DFT) via the Vienna ab initio Simulation Package (VASP).[98] The Perdew-Burke-Ernzerhof (PBE) exchange-correlation functional within the generalized gradient approximation (GGA) was performed, alongside the projected augmented wave (PAW) method.[99] The PBE + $U_{\mathrm{eff}}$ approach[100] with effective

Hubbard $U_{eff}$ = 4.0 eV was implemented for V-3*d* orbitals, which was consistent with that in previous studies.[61,62,101,102]

For spin transport properties, calculations were performed using the QuantumWise Atomistix ToolKit (ATK) package, which combines DFT with the non-equilibrium Green's function (NEGF) method.[103]

Computational details are presented in Supporting Information. These computational methods and parameters have been applied in our previous investigations.[3,8,19,104]

**Supporting Information**

Supporting Information is available from the Wiley Online Library or from the author.

**Acknowledgements**

We sincerely thank Prof. Tian Qian from Institute of Physics, Chinese Academy of Sciences, for his beneficial discussion on experimental altermagnetism of $V_2Se_2O$ family, and we also thank Asst. Prof. Zhuang Ma from Zhoukou Normal University, for his helpful input on multiferroics, particularly ferroelasticity.

Guoying Gao acknowledges support from the National Natural Science Foundation of China (Grant No. 12174127). Guangqian Ding acknowledges support from the National Natural Science Foundation of China (Grant No. 12374002). Guangxin Ni acknowledges support from the U.S. Department of Energy under award DE-SC0022022, National Science Foundation under award DMR-2145074, and ACS-DNI (PRF# 66465-DNI10).

**Conflict of Interest**

The authors declare no conflict of interest.

**Data Availability Statement**

The data that support the findings of this study are available from the corresponding author upon reasonable request.

## Figures and Tables

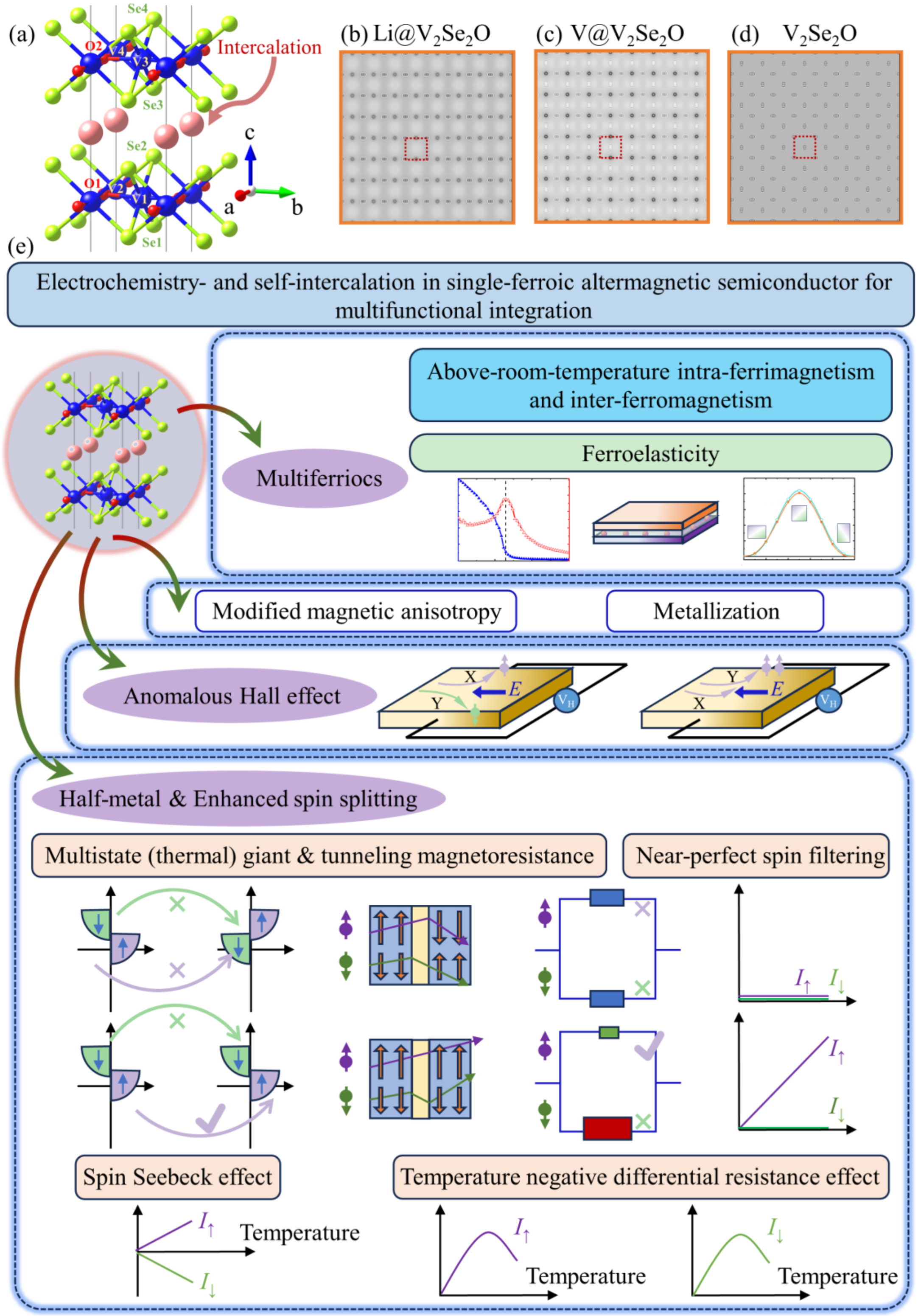

**Figure 1.** The crystal structure of intercalated $V_2Se_2O$ bilayer (a). Simulated scanning tunneling microscope (STM) image of $V_2Se_2O$ bilayer with Li (b), V (c), and no (d) intercalations on the (001) plane at the sample bias voltage ($V_s$) of 0.5 V, where the red dotted box is a unit cell. Schematic illustrations of multifunctional integration for intercalated $V_2Se_2O$ bilayers (e).

**Table 1.** In-plane lattice constants, reversible strain, inter Se-Se distances, and V-V distances. The V-V distance $d_1$, $d_{2x}$, $d_{2y}$ and $d_{int}$ correspond to the considered parameters of magnetic exchange interactions, and the $d_{int}$ is chosen as the distance between the inter-center V atoms.

| | In-plane lattice constant (Å) | | Reversible strain (%) | Inter Se-Se distance (Å) | V-V distance (Å) | | | | |
|---|---|---|---|---|---|---|---|---|---|
| | $a$ | $b$ | $\varepsilon$ | $d_{Se}$ | $d_1$ | $d_{11}$ | $d_{2x}$ | $d_{2y}$ | $d_{int}$ |
| Li@$V_2Se_2O$ | 4.058 | 4.033 | 0.616 | 3.526 | 2.861 | 5.722 | 4.058 | 4.033 | 7.133 |
| V@$V_2Se_2O$ | 4.100 | 4.020 | 1.990 | 3.497 | 2.871 | 5.742 | 4.100 | 4.020 | 7.223 |
| $V_2Se_2O$ | 4.040 | 4.040 | 0.000 | 3.707 | 2.857 | 5.714 | 4.040 | 4.040 | 7.147 |

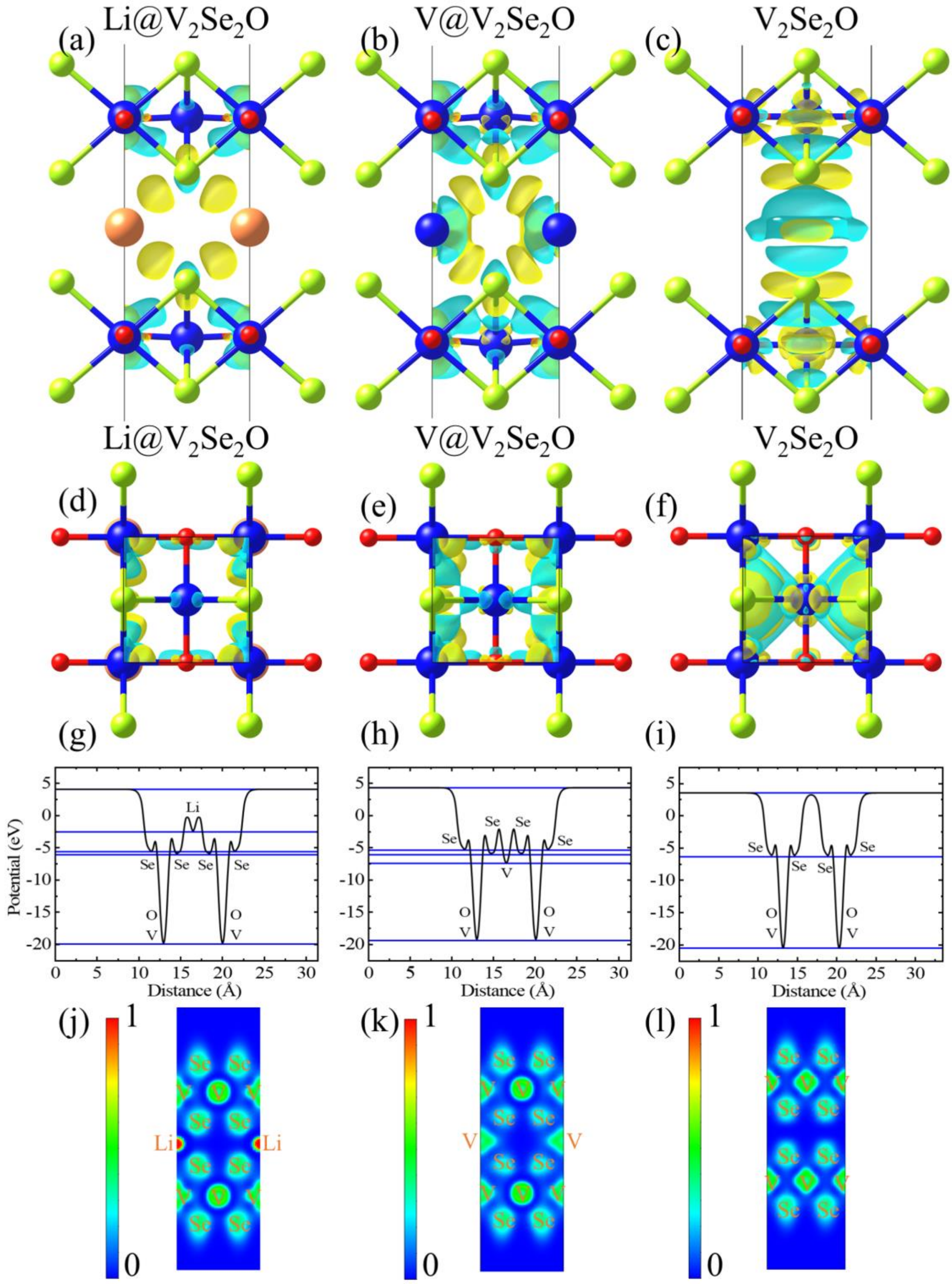

**Figure 2.** The side (a-c) and top (d-f) views for the charge density difference of the $V_2Se_2O$ bilayer with Li-, V-, and no intercalations, in which the isosurface value is 0.003 $e$ bohr$^{-3}$. The plane-averaged electrostatic potentials along the $z$-direction (g-i) and electron localization functions (ELFs) in the (110) plane (j-l) of Li-intercalated, V-intercalated and pristine $V_2Se_2O$ bilayer.

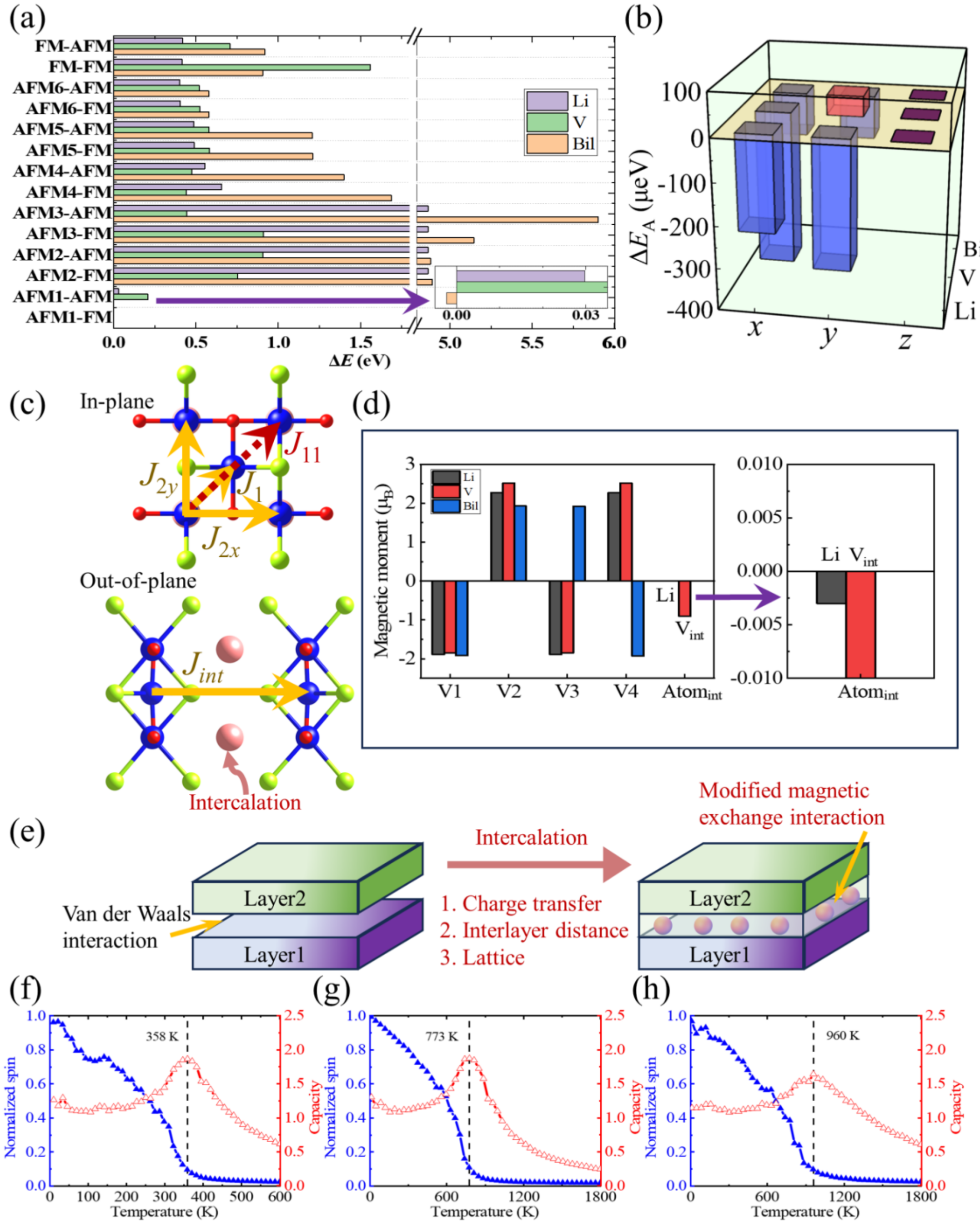

**Figure 3.** The energy differences ($\Delta E$s) with reference to the AFM1-FM states (a), which represent the intralayer AFM1 and interlayer FM configurations. Based on the magnetic ground configurations, the anisotropy energy differences ($\Delta E_A$s) with reference to the states with magnetization axes along *z*-directions (b). The in-plane and out-of-plane exchange couplings (c). The main atomic magnetic moments (d). The mentioned energies are among the unit cell, and Li, V and Bil in (a,b,d) succinctly represent Li-intercalated, V-intercalated and pristine $V_2Se_2O$ bilayer, respectively. Schematic illustrations for modified magnetic exchange interaction in the altermagnetic bilayer system through intercalation engineering (e). The variations of the normalized spin operator (blue solid) and specific heat capacity (red hollowed) with the temperature for $V_2Se_2O$ bilayer with Li (f), V (g), and no (h) intercalations.

**Table 2.** The parameters of magnetic exchange interactions of $V_2Se_2O$ bilayer with and without intercalations. $J_1$, $J_{2x}$, $J_{2y}$, $J_{11}$, and $J_{int}$ in the crystal lattice are shown in Figure 3c.

| Parameters (meV f.u.$^{-1}$) | $J_1$ | $J_{2x}$ | $J_{2y}$ | $J_{11}$ | $J_{int}$ |
|---|---|---|---|---|---|
| Li@$V_2Se_2O$ | -25.952 | 17.347 | 38.771 | -12.128 | 7.423 |
| V@$V_2Se_2O$ | -97.464 | -28.417 | -13.999 | 3.326 | 51.728 |
| $V_2Se_2O$ | -56.673 | -1.489 | 155.909 | -60.723 | -0.584 |

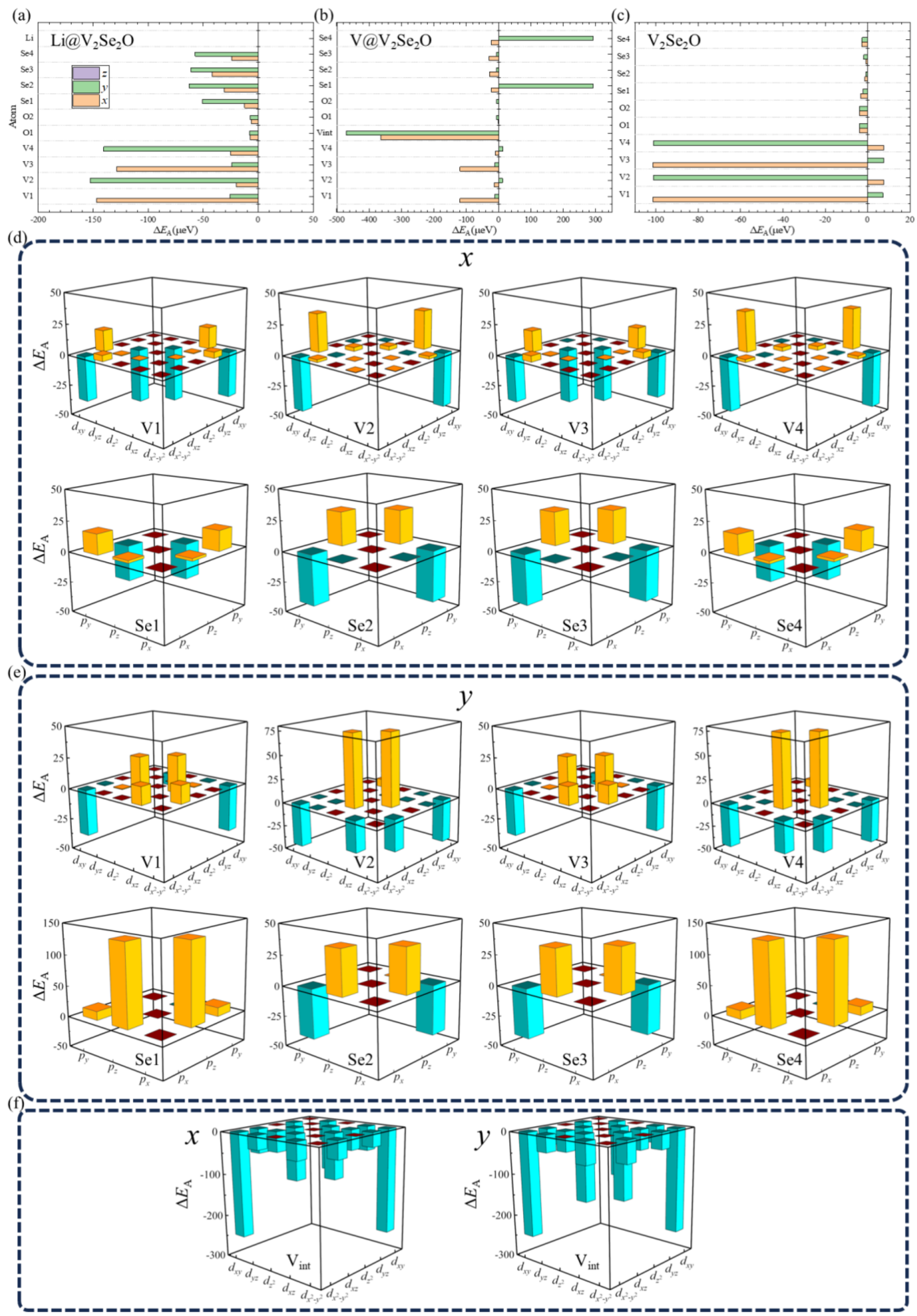

**Figure 4.** The atom-resolved $\Delta E_A$s for $V_2Se_2O$ bilayer with Li (a), V (b), and no (c) intercalations. The V-*d*- and Se-*p*-orbital-resolved $\Delta E_A$s for V-intercalated $V_2Se_2O$ bilayer with magnetization axis along *x*- (d) and *y*- (e) directions, and the contribution from intercalated V ($V_{int}$) (f). The $\Delta E_A$s are with reference to the states with magnetization axis along the *z*-direction.

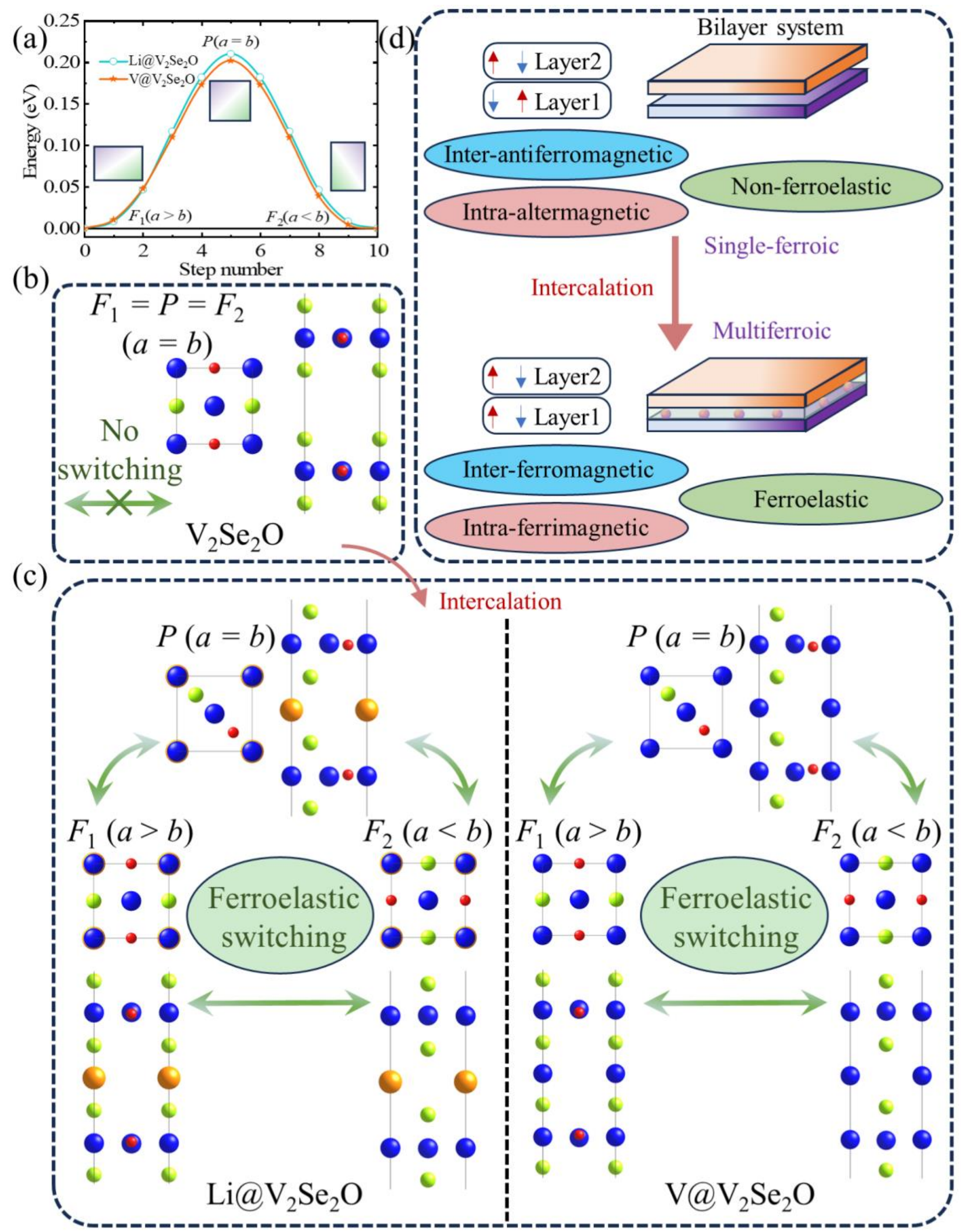

**Figure 5.** Energy barriers of ferroelastic (FC) transition for Li- and V-intercalated $V_2Se_2O$ bilayer (a). No switching for pristine $V_2Se_2O$ bilayer (b), and FC switching for the $V_2Se_2O$ bilayer with Li (c) and V (d) intercalations. Schematic illustration for the transition from single-ferroic to multiferroic coupling in the altermagnetic bilayer through intercalation engineering (d).

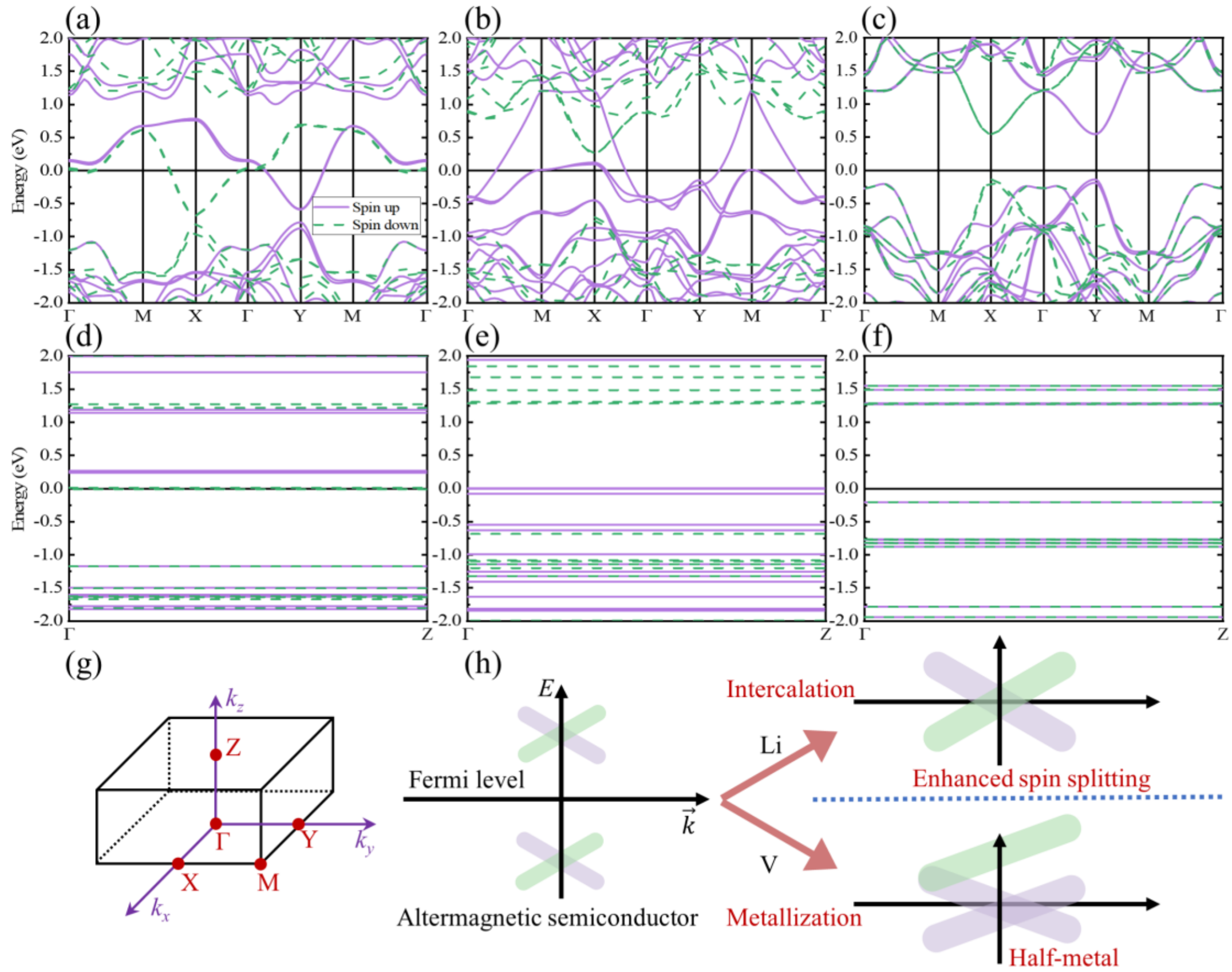

**Figure 6.** The spin-resolved band structures for $V_2Se_2O$ bilayer with Li, V and no intercalations along the Γ—M—X—Γ—Y—M—Γ (a-c) and Γ—Z (d-f) pathways. The high-symmetry points in the Brillouin zone (g). Schematic illustration for the modified electronic structures of the altermagnetic semiconductor through intercalation engineering (h). Li- and V-intercalations induce metallicity with enhanced spin splitting and half-metallicity, respectively.

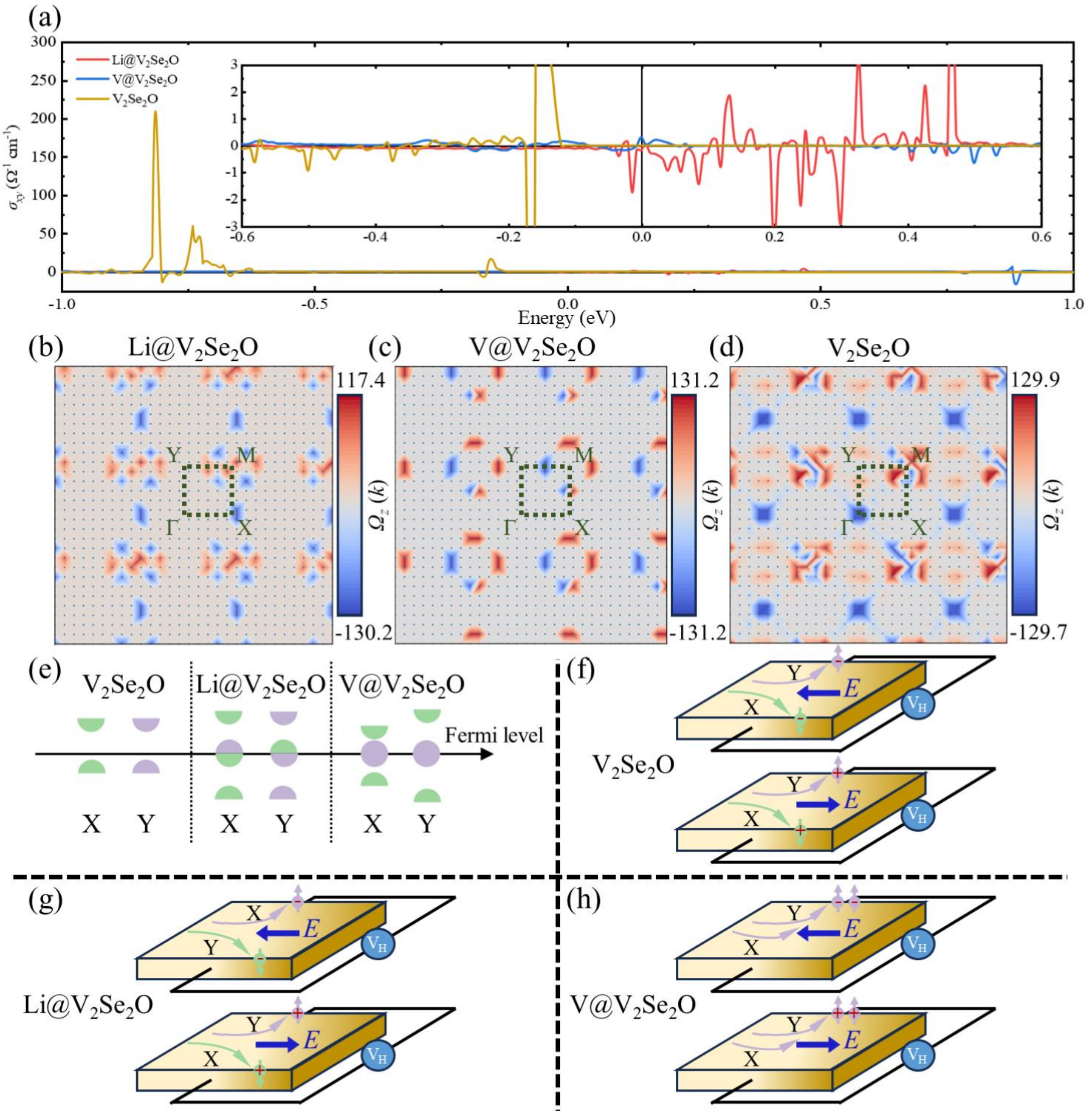

**Figure 7.** Energy-dependent anomalous Hall conductivity (AHC) (a), and corresponding momentum-dependent Berry curvature in 2D Brillouin zone of $V_2Se_2O$ bilayer with Li (b), V (c), and no (d) intercalations. Schematic illustration for the modified momentum- and spin-dependent electronic states around the Fermi level of the altermagnetic system through intercalation engineering (e) and corresponding anomalous Hall effect devices (f-h).

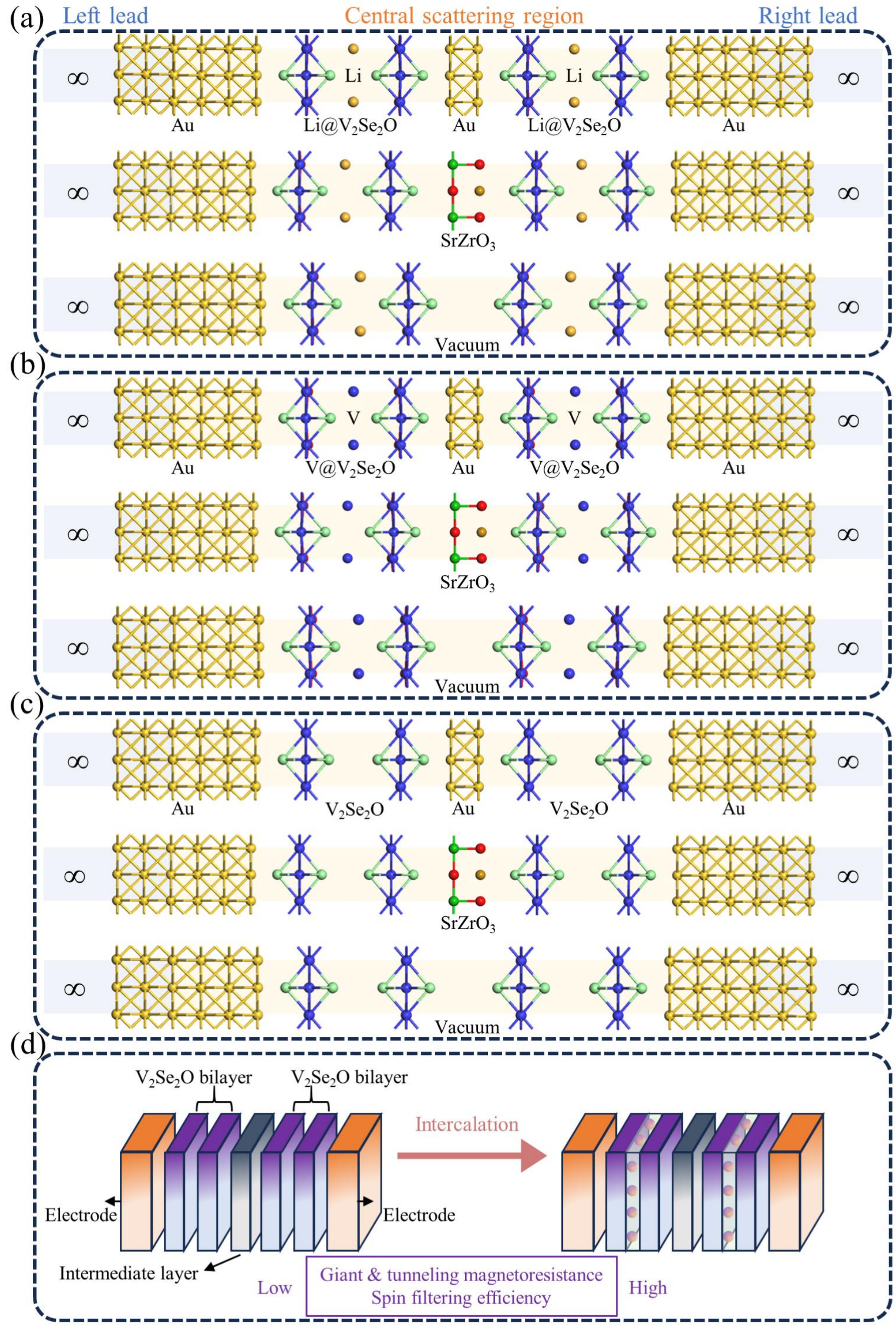

**Figure 8.** The $V_2Se_2O$-based magnetoresistance (MR) devices with an electrode of Au and intermediate layers of Au, $SrZrO_3$ or vacuum, in which $V_2Se_2O$ bilayer with Li (a), V (b) and no (c) intercalations. Schematic illustration for the enhanced spin transport performance of altermagnet through intercalation engineering (d).

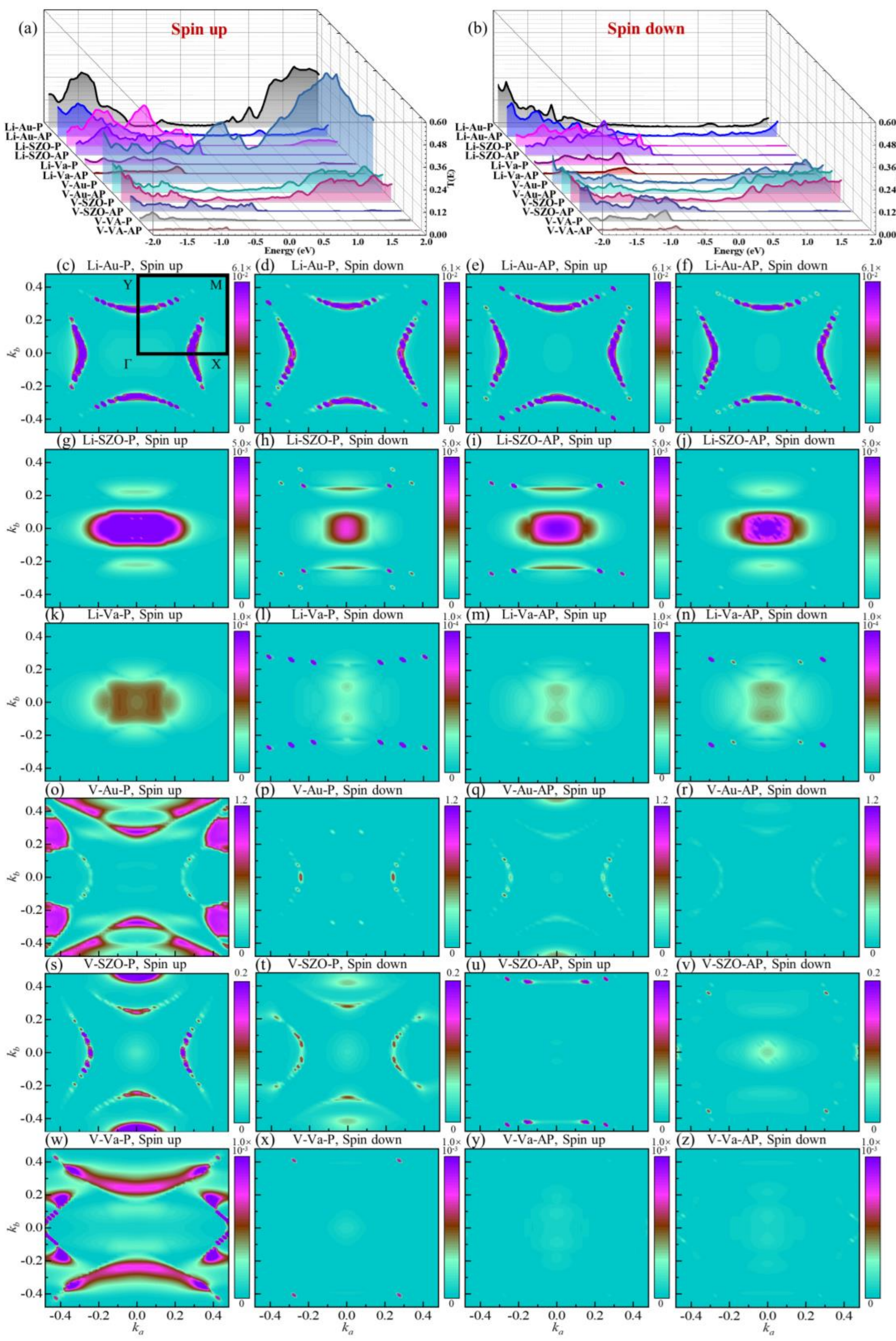

**Figure 9.** The transmission coefficients as a function of energy in spin-up (a) and down (b) channels. The $\vec{k}_{//}$- and spin-resolved transmission spectrum in the 2D Brillouin zone across $V_2Se_2O$-based MR devices (c-z).

**Table 3.** Spin-dependent electron transmission $T_{\uparrow}$ and $T_{\downarrow}$, spin filtering efficiency $\eta$, and MR ratios across $V_2Se_2O$-based MR devices.

| | P | | | | AP | | | | MR (%) |
|---|---|---|---|---|---|---|---|---|---|
| | $T_{\uparrow}$ | $T_{\downarrow}$ | $T_{\mathrm{tot}}$ | $\eta$ (%) | $T_{\uparrow}$ | $T_{\downarrow}$ | $T_{\mathrm{tot}}$ | $\eta$ (%) | |
| Li-Au | $8.97 \times 10^{-3}$ | $7.52 \times 10^{-3}$ | $1.65 \times 10^{-2}$ | 9 | $9.07 \times 10^{-3}$ | $7.07 \times 10^{-3}$ | $1.61 \times 10^{-2}$ | 12 | 2 |
| Li-SZO | $6.61 \times 10^{-4}$ | $2.68 \times 10^{-4}$ | $9.30 \times 10^{-4}$ | 42 | $4.17 \times 10^{-4}$ | $3.85 \times 10^{-4}$ | $8.02 \times 10^{-4}$ | 4 | 16 |
| Li-Va | $7.08 \times 10^{-6}$ | $6.95 \times 10^{-5}$ | $7.66 \times 10^{-5}$ | 82 | $3.70 \times 10^{-6}$ | $5.29 \times 10^{-6}$ | $8.99 \times 10^{-6}$ | 18 | 677 |
| V-Au | $2.55 \times 10^{-1}$ | $4.22 \times 10^{-3}$ | $2.59 \times 10^{-1}$ | 97 | $1.80 \times 10^{-2}$ | $8.54 \times 10^{-3}$ | $2.66 \times 10^{-2}$ | 36 | 877 |
| V-SZO | $2.06 \times 10^{-2}$ | $2.40 \times 10^{-4}$ | $2.08 \times 10^{-2}$ | 98 | $2.14 \times 10^{-3}$ | $4.09 \times 10^{-4}$ | $2.55 \times 10^{-3}$ | 68 | 716 |
| V-Va | $2.32 \times 10^{-4}$ | $4.17 \times 10^{-6}$ | $2.37 \times 10^{-4}$ | 96 | $8.28 \times 10^{-6}$ | $1.00 \times 10^{-5}$ | $1.83 \times 10^{-5}$ | 9 | 1194 |
| Bil-Au | $8.10 \times 10^{-3}$ | $7.37 \times 10^{-3}$ | $1.55 \times 10^{-2}$ | 5 | $6.32 \times 10^{-3}$ | $7.60 \times 10^{-3}$ | $1.39 \times 10^{-2}$ | 9 | 11 |
| Bil-SZO | $6.22 \times 10^{-4}$ | $3.86 \times 10^{-4}$ | $1.01 \times 10^{-3}$ | 23 | $4.38 \times 10^{-4}$ | $6.04 \times 10^{-4}$ | $1.04 \times 10^{-3}$ | 16 | -3 |
| Bil-Va | $1.38 \times 10^{-5}$ | $1.13 \times 10^{-5}$ | $2.50 \times 10^{-5}$ | 10 | $1.14 \times 10^{-5}$ | $1.18 \times 10^{-5}$ | $2.32 \times 10^{-5}$ | 2 | 8 |

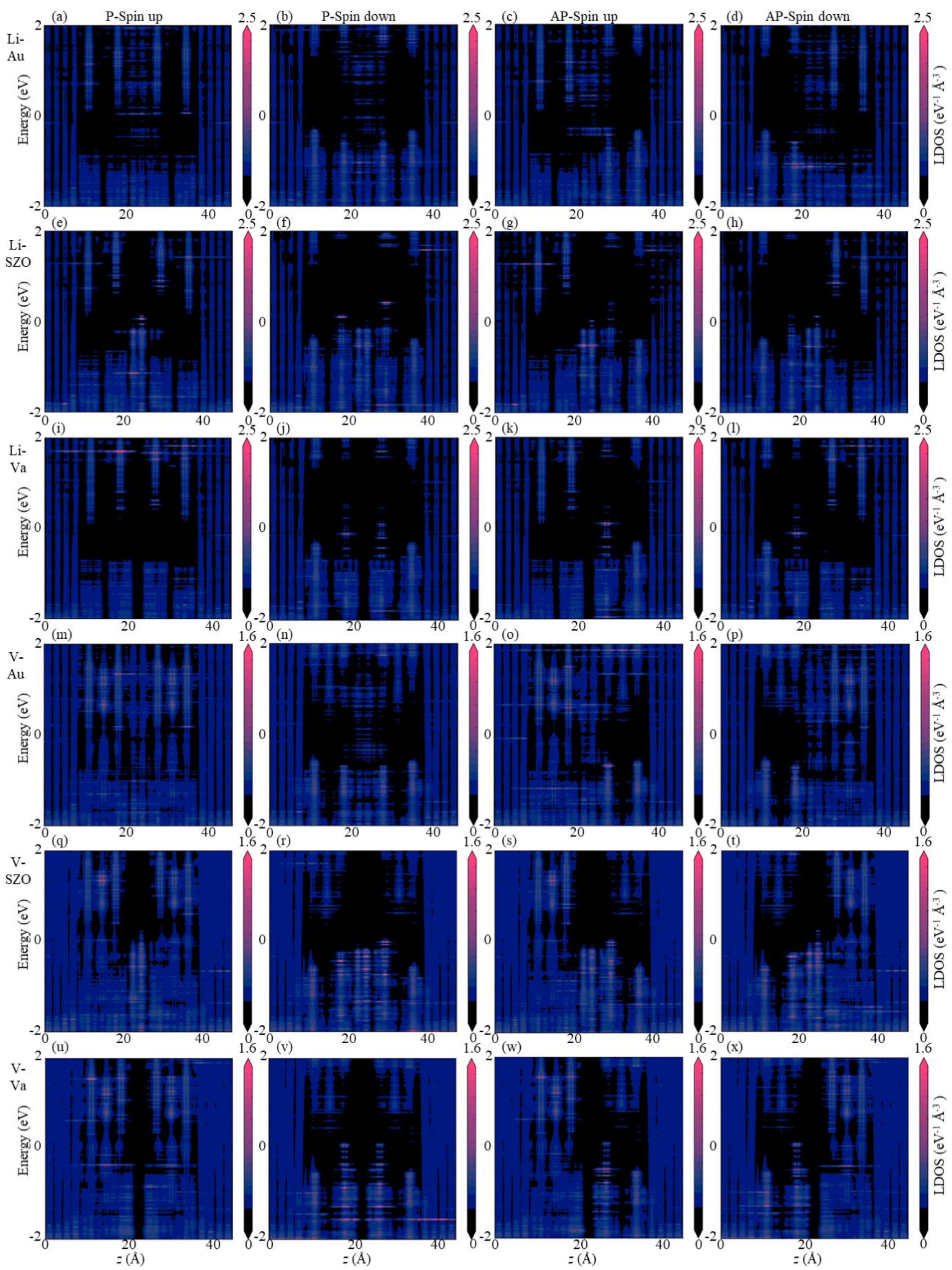

**Figure 10.** The spin-resolved local density of states (LDOS) across $V_2Se_2O$-based MR devices (a-x).

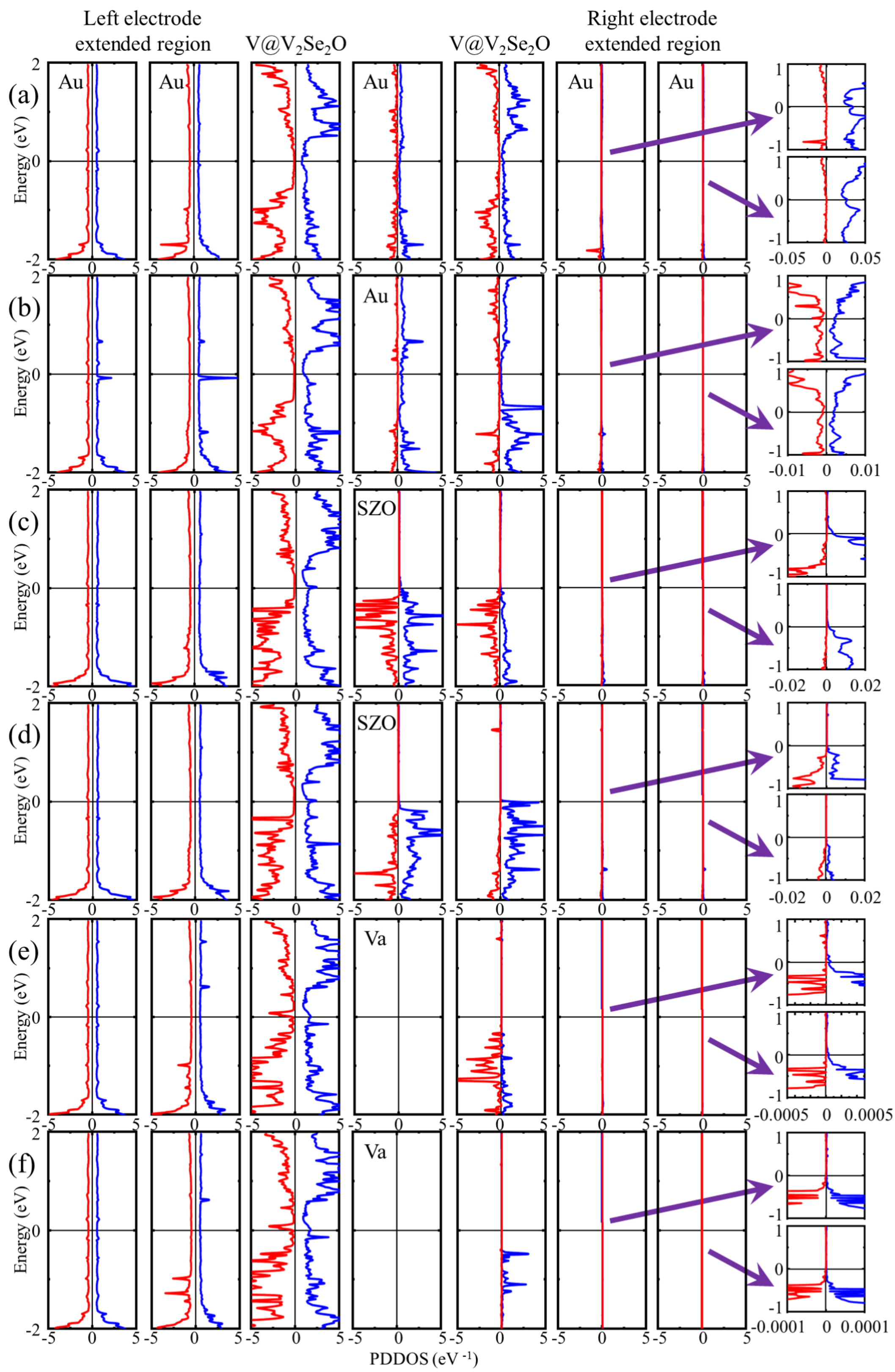

**Figure 11.** The spin- and layer-resolved projected device density of states (PDDOS) across $V_2Se_2O$-based MR devices, which is constructed by the parallel and antiparallel V-intercalated $V_2Se_2O$ bilayers and the intermediate layer of Au (a,b), $SrZrO_3$ (c,d) and vacuum (e,f). The source of contribution is set as the left Au electrode.

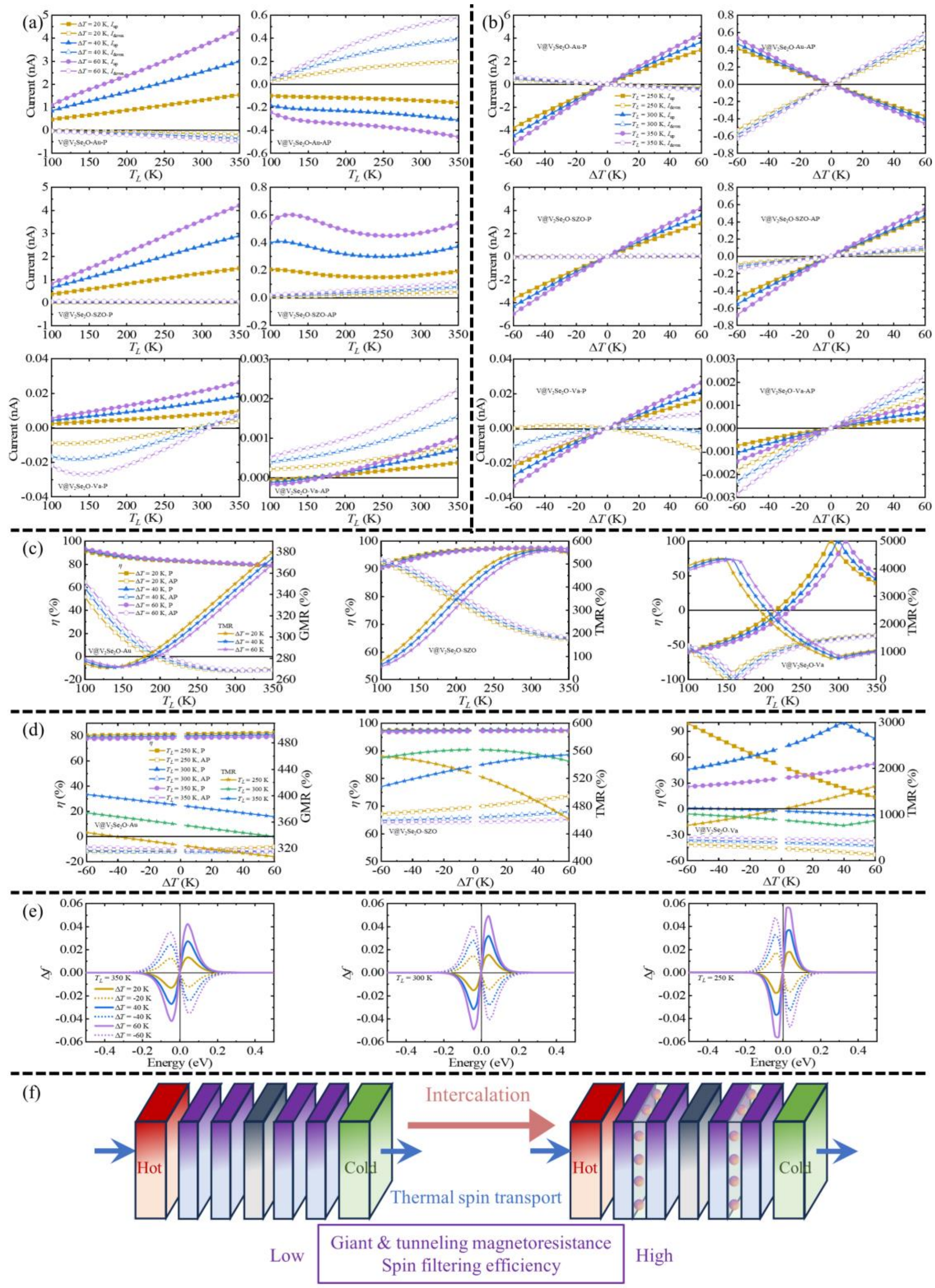

**Figure 12.** Thermal spin transport performance of V-intercalated $V_2Se_2O$ bilayer, Au/SZO/Va stands for the intermediate layer of Au, $SrZrO_3$ and vacuum, and P and AP stand for the parallel and antiparallel configurations, respectively. The thermal spin-dependent current, versus $T_L$ and $\Delta T$ for different $\Delta T$ and $T_L$ (a,b), the thermal spin filtering efficiency $\eta$ and TMR versus $T_L$ for different $\Delta T$ (c) and versus $\Delta T$ for different $T_L$ (d). The Fermi-Dirac distribution difference $\Delta f$ for different $T_L$ and $\Delta T$ (e). Schematic illustration for the enhanced thermal spin transport performance of altermagnet through intercalation engineering (f).

**Table 4.** Spin-dependent current $I_\uparrow$ and $I_\downarrow$, spin filtering efficiency $\eta$, and MR ratios across $V_2Se_2O$-based MR devices at $T_L$ = 300 K and $\Delta T$ = 20 K.

| | P | | | | AP | | | | MR (%) |
|---|---|---|---|---|---|---|---|---|---|
| | $I_\uparrow$ (nA) | $I_\downarrow$ (nA) | $I_{tot}$ (nA) | $\eta$ (%) | $I_\uparrow$ (nA) | $I_\downarrow$ (nA) | $I_{tot}$ (nA) | $\eta$ (%) | |
| Li-Au | $1.41 \times 10^{-1}$ | $1.40 \times 10^{-1}$ | $2.81 \times 10^{-1}$ | 0 | $8.27 \times 10^{-2}$ | $1.14 \times 10^{-1}$ | $1.97 \times 10^{-1}$ | -16 | 43 |
| Li-SZO | $1.03 \times 10^{-1}$ | $8.45 \times 10^{-2}$ | $1.87 \times 10^{-1}$ | 10 | $8.05 \times 10^{-2}$ | $9.76 \times 10^{-2}$ | $1.78 \times 10^{-1}$ | -10 | 5 |
| Li-Va | $1.90 \times 10^{-4}$ | $7.82 \times 10^{-3}$ | $8.01 \times 10^{-3}$ | -95 | $1.56 \times 10^{-4}$ | $1.67 \times 10^{-4}$ | $3.23 \times 10^{-4}$ | -3 | 2385 |
| V-Au | 1.31 | $-1.45 \times 10^{-1}$ | 1.45 | 80 | $-1.42 \times 10^{-1}$ | $1.81 \times 10^{-1}$ | $3.23 \times 10^{-1}$ | -12 | 350 |
| V-SZO | 1.27 | $1.63 \times 10^{-2}$ | 1.29 | 97 | $1.63 \times 10^{-1}$ | $3.27 \times 10^{-2}$ | $1.95 \times 10^{-1}$ | 66 | 559 |
| V-Va | $7.71 \times 10^{-3}$ | $6.92 \times 10^{-4}$ | $8.40 \times 10^{-3}$ | 84 | $2.73 \times 10^{-4}$ | $6.38 \times 10^{-4}$ | $9.12 \times 10^{-4}$ | -40 | 821 |
| Bil-Au | $1.47 \times 10^{-1}$ | $3.71 \times 10^{-2}$ | $1.84 \times 10^{-1}$ | 60 | $2.34 \times 10^{-2}$ | $1.82 \times 10^{-1}$ | $2.05 \times 10^{-1}$ | -77 | -10 |
| Bil-SZO | $1.12 \times 10^{-1}$ | $8.35 \times 10^{-2}$ | $1.95 \times 10^{-1}$ | 15 | $9.34 \times 10^{-2}$ | $1.01 \times 10^{-1}$ | $1.95 \times 10^{-1}$ | -4 | 4 |
| Bil-Va | $8.21 \times 10^{-4}$ | $7.35 \times 10^{-4}$ | $1.56 \times 10^{-3}$ | 6 | $7.18 \times 10^{-4}$ | $7.62 \times 10^{-4}$ | $1.48 \times 10^{-3}$ | -3 | 5 |